\documentclass[preprint,onecolumn,nofootinbib]{revtex4}
\pdfoutput=1
\usepackage[colorlinks=true,linkcolor=blue,urlcolor=blue,filecolor=black,citecolor=red,pdfstartview=FitV,pdftitle={},pdfsubject={},pdfkeywords={},pdfpagemode=None,bookmarksopen=true]{hyperref}
\usepackage{graphicx}%Include figure files
\usepackage{subfigure}
\usepackage{amsthm,amsmath,amssymb}
\usepackage{url}
\usepackage{array}
\usepackage{float}
\usepackage{appendix}
\usepackage{mathrsfs}%包含命令“\mathscr{}”（花体字母）
\usepackage{amsmath}
\usepackage{mathtools}%amsmath宏包的拓展，可以使用更美观的矢量箭头符号
\usepackage{amsfonts}
\usepackage{amssymb,ulem}
\usepackage{color,xcolor}%
\usepackage{CJK}

\usepackage{dcolumn}% Align table columns on decimal pointl

\newcommand{\f}{\begin{equation}}
\newcommand{\ff}{\end{equation}}
\newcommand{\fa}{\begin{eqnarray}}
\newcommand{\ffa}{\end{eqnarray}}
\begin{document}

\title{Non-singular cosmologies matching regular black holes}

\author{Shulan Li$^{1}$}
\thanks{shulanli.yzu@gmail.com}
\author{Jian-Pin Wu$^{2}$}
\thanks{jianpinwu@yzu.edu.cn, corresponding author} 
\author{Xian-Hui Ge$^{1}$}
\thanks{gexh@shu.edu.cn, corresponding author}
\affiliation{
  $^{1}$ \mbox{Department of Physics, Shanghai University, Shanghai 200444, China}\\
  $^{2}$ \mbox{Center for Gravitation and Cosmology, College of Physical Science and Technology,} \mbox{Yangzhou University, Yangzhou 225009, China}
  }

%====================================================
\begin{abstract}
We construct a new non-singular cosmological model matched to a Minkowski-core regular black hole by means of a modified Oppenheimer--Snyder framework.
Its dynamics in both dust-only and scalar-field scenarios is studied and compared with that of two other non-singular models as well as the classical standard cosmology.
The results show that, although all three non-singular cosmologies share identical late-time behavior and allow for a natural embedding of inflation in the scalar-field setting, they exhibit qualitatively distinct non-singular features at very early times.
For the new cosmology specifically, we also investigate the scenarios of other cosmological fluids and, more generally, of multiple coexisting fluids.
This cosmological model is found to manifest an intriguing symmetry between its asymptotic past and future: without a cosmological constant both limits are asymptotically Minkowski, whereas with a positive one both are asymptotically de Sitter.
Furthermore, a dynamical stability analysis reveals that the asymptotic-past state of this model is a robust attractor, thereby requiring no extreme fine-tuning of the initial conditions.
\end{abstract}

%====================================================
\maketitle
\tableofcontents

%====================================================
\section{Introduction}
\label{introduction}
%----------------------------------------------------
General relativity (GR) has convincingly elucidated that the nature of the gravitational interactions is the curvature of spacetime.
This theory has not only predicted the existence and formation of black holes (BHs) but also provided the foundation of modern cosmology.
Nevertheless, Penrose and Hawking's singularity theorem points out the occurrence of spacetime singularities in both the final stage of the gravitational collapse of massive celestial bodies and the initial phase of the expanding universe.
This reveals the breakdown of classical GR in extremely high spacetime curvature regions at and below Planckian scales, indicating that quantum gravity effects may play a significant role there.

To avoid the singularity inside BHs, a multitude of regular (i.e., non-singular) BH models have been proposed to date.
The concept of singularity usually involves the incompleteness of null and timelike geodesics \cite{Hawking1973LargeScale, Wald:1984rg}, as well as the divergence of spacetime curvature invariants \cite{weinberg1972gravitation, markov1982limiting, Frolov:1988vj, Frolov:2016pav, Chamseddine:2016ktu, Dymnikova:1992ux, Ayon-Beato:1998hmi, Bronnikov:2000vy}, which are closely related but not completely equivalent.
Some of the regular BH models are based on the candidate theories of quantum gravity, such as loop quantum gravity (LQG) \cite{Ashtekar:2005qt, Modesto:2005zm, Campiglia:2007pr, Boehmer:2007ket, Chiou:2008nm, Gambini:2013ooa, Gambini:2013hna, Corichi:2015xia, Olmedo:2017lvt, Bianchi:2018mml, Bojowald:2018xxu, BenAchour:2018khr, Morales-Tecotl:2018ugi, Assanioussi:2019twp, Arruga:2019kyd, BenAchour:2020gon, Bojowald:2020dkb, Blanchette:2020kkk, Assanioussi:2020ezr, Gan:2022oiy, Peltola:2008pa, Peltola:2009jm, Modesto:2008im, Ashtekar:2018lag, Ashtekar:2018cay, Gambini:2020nsf, Bodendorfer:2019cyv, Bodendorfer:2019nvy, Kelly:2020uwj, Parvizi:2021ekr, Lewandowski:2022zce, Giesel:2022rxi, Alonso-Bardaji:2021yls, Alonso-Bardaji:2022ear, Zhang:2024khj, Zhang:2024ney} and asymptotically safe gravity \cite{Bonanno:2000ep, Koch:2014cqa}; while others arose at the phenomenological level---often involving the violation of some energy conditions within the framework of classical GR or other modified gravity theories \cite{Bardeen:1968, Hayward:2005gi, Ansoldi:2008jw, Frolov:2014jva, Neves:2014aba, Dymnikova:2015yma, Maluf:2018lyu, LiXiang:2013sza, Culetu:2013fsa, Culetu:2014lca, Balart:2014cga, Rodrigues:2015ayd, Simpson:2019mud, Ghosh:2014pba, Ghosh:2018bxg, Ling:2021olm, Bronnikov:2000vy, Ayon-Beato:2000mjt, Pedraza:2020uuy, Fan:2016hvf, Bronnikov:2005gm, Bronnikov:2021uta, Bokulic:2022cyk, Canate:2022zst, Cisterna:2020rkc, Dymnikova:1992ux, Nicolini:2005vd, Nicolini:2008aj, Spallucci:2008ez, Nicolini:2009gw, Balakin:2006gv, Balakin:2016mnn, Roupas:2022gee, Simpson:2018tsi, Carballo-Rubio:2019fnb, Carballo-Rubio:2022kad, Capozziello:2024ucm, Capozziello:2025wwl, DeBianchi:2025bgn, Parvez:2025wtq}.
For the latter approach, according to the asymptotic behavior near the center of the BH, there are two typical cases: de Sitter (dS) core and Minkowski core.
Regular BHs with dS cores \cite{Bardeen:1968, Ayon-Beato:2000mjt, Hayward:2005gi, Pedraza:2020uuy, Ansoldi:2008jw, Frolov:2014jva, Neves:2014aba, Dymnikova:2015yma, Maluf:2018lyu} asymptotically approach the dS spacetime near the center, including the well-known Bardeen, Hayward, and Frolov BHs.
In contrast, regular BHs with Minkowski cores \cite{LiXiang:2013sza, Culetu:2013fsa, Culetu:2014lca, Balart:2014cga, Rodrigues:2015ayd, Simpson:2019mud, Ghosh:2014pba, Ghosh:2018bxg, Ling:2021olm, Chen:2026hmt} tend to the flat Minkowski spacetime near the center, which are characterized by an exponentially suppressing gravitational potential.
For more detailed analysis and discussion on regular BHs, one can refer to the reviews and related literature \cite{Ansoldi:2008jw, Nicolini:2008aj, Torres:2022twv, Giacchini:2021pmr, Zhou:2022yio, Sebastiani:2022wbz, bambi2023regular, Lan:2023cvz}.

Oppenheimer--Snyder (OS) model \cite{Oppenheimer:1939ue}, the gravitational collapse model of a spherically symmetric homogeneous dust ball, is not only a simple yet effective framework to understand and investigate the dynamical formation of BHs and their singularities, but also an especially suitable arena to explore the intrinsic connection between the central singularity problem in BHs and the initial singularity problem in cosmology.
Within the GR framework, the collapsing dust ball is described by the Friedmann--Robertson--Walker (FRW) metric in the standard cosmology in its interior and the Schwarzschild metric in its exterior, which join smoothly at the boundary surface via the Darmois--Israel (DI) junction conditions \cite{Darmois1927, Israel1966}.
This implies that the OS model can naturally serve as a bridge between BH spacetimes and cosmological spacetimes.
Furthermore, related studies that explore the relationship between these two types of spacetimes also include, but are not limited to, the formation of primordial BHs in the early universe \cite{Konoplich:1999qq, Khlopov:2000js, Khlopov:2008qy, Dymnikova:2015yma} and the dynamics of BHs embedded in the accelerating universe \cite{Nojiri:2025heh}.
In recent years, a variety of quantum modified versions for the OS collapse model have been widely studied \cite{Maier:2009xp, BenAchour:2020bdt, BenAchour:2020gon, Piechocki:2020bfo, Kelly:2020lec, Munch:2020czs, Schmitz:2020vdr, Husain:2021ojz, Malafarina:2022oka, Lewandowski:2022zce, Yang:2022btw, Han:2023wxg, Bobula:2023kbo, Bonanno:2023rzk, Giesel:2023tsj, Giesel:2023hys, Cafaro:2024vrw, Bobula:2024jlh, Giesel:2024mps, Harada:2025cwd, Bueno:2025gjg}.
In particular, a quantum OS model utilizing the interior-to-exterior approach was recently put forward in Refs.~\cite{Lewandowski:2022zce, Yang:2022btw}, where the interior quantum-deformed geometry from loop quantum cosmology (LQC) \cite{Ashtekar:2006rx, Ashtekar:2006uz, Ashtekar:2006wn} is carried over to the exterior, thereby leading to a quantum BH model.
More recently, the author of Ref.~\cite{Bobula:2024jlh} adopted the Hayward metric \cite{Hayward:2005gi} in place of the Schwarzschild metric to describe the static, spherically symmetric vacuum exterior region in his modified OS model, and then derived and discussed the matching interior cosmological behavior, which turned out to be non-singular.
Considering that the aforementioned work \cite{Bobula:2024jlh} investigated the cosmology corresponding to the Hayward BH model, a typical example of regular BHs with dS cores, we are now interested in a parallel question: what about the cosmology when matched with a regular BH possessing a Minkowski core?
To this end, in the present work, we will set up a new modified OS scenario where the exterior vacuum region is described by the metric of the Minkowski-core regular BH proposed in Ref.~\cite{Ling:2021olm}, so as to derive and analyze the corresponding interior cosmological dynamics.
The resulting cosmological model will then be compared with the following three representative models: the cosmology matched to the Hayward BH given in Ref.~\cite{Bobula:2024jlh}, LQC, and the standard cosmology within classical GR.

The rest of this paper is organized as follows.
In Sec.~\ref{1}, we review the OS collapse scenario and propose a new non-singular cosmological model matched to a Minkowski-core regular BH based on this framework.
Building upon this, Sec.~\ref{2} presents comparative studies on the dynamical evolution of the obtained cosmological model with LQC, the Hayward cosmology, and the classical standard cosmology, including both the dust-only scenario directly derived from the OS setup and an alternative scenario involving a massive scalar field; for the new cosmology specifically, it further examines the scenarios of other cosmological fluids and the more realistic multi-fluid scenario, and evaluates the robustness of its asymptotic past within a dynamical systems framework.
Finally, Sec.~\ref{conclusion} summarizes the main findings and provides a brief discussion.
Moreover, a detailed derivation of the interior-exterior matching in the OS model by means of the DI junction conditions is provided in Appendix~\ref{AppendixA}, and the two matching pairs of regular BH and non-singular cosmological models for comparison are briefly introduced in Appendices~\ref{AppendixB} and \ref{AppendixC}.

We would like to remind our readers that the Planck unit system is used throughout this paper, with $ c = G = \hbar = 1 $.
Here, $c$ stands for the speed of light, $G$ represents the gravitational constant, and $\hbar$ denotes the reduced Planck constant.

%====================================================
\section{New non-singular cosmology via the Oppenheimer--Snyder scenario}
\label{1} %"the 1st section of the main body"
%----------------------------------------------------
In this section, we first review the OS dust collapse model in the context of classical GR, after which a modified version of this framework is applied to present a new non-singular cosmological model that matches the Minkowski-core regular BH model presented in Ref.~\cite{Ling:2021olm}.

%----------------------------------------------------
\subsection{Oppenheimer--Snyder dust collapse model}
\label{1.1} %"1.1 of the main body"
%----------------------------------------------------
In the OS collapse model \cite{Oppenheimer:1939ue}, the timelike boundary hypersurface $\varSigma = {{\cal M}^ - } \cap {{\cal M}^ + }$ divides the spacetime manifold ${\cal M}$ into the interior region ${{\cal M} ^-}$ and the exterior region ${{\cal M} ^+}$.
It is assumed that the interior region ${{\cal M} ^-}$ is constructed by a spherically symmetric collapsing dust ball with uniform density and zero pressure.
The spacetime geometry of this region is equivalent to that of a collapsing dust-only universe, described by a spatially flat FRW metric\footnote{In cosmology, the spatial geometry of the universe can be classified into three cases: a 3-dimensional sphere, a 3-dimensional Euclidean space, and a 3-dimensional hyperboloid, corresponding to what are commonly referred to as the closed, flat, and open universe models, respectively. In this work, we focus exclusively on the flat case.}:
\begin{equation}
\label{IM} %"the Interior Metric"
\mathrm{d}s_ - ^2 \equiv {g^{-}_{\mu \nu}} {\mathrm{d} x^{\mu}} {\mathrm{d} x^{\nu}} = - {\mathrm{d}t} ^2 + {a\!\left( t \right)}^2\left( {{\mathrm{d}{r}}^2 + {{r}^2}{\mathrm{d}\Omega} ^2} \right)\;.
\end{equation}
Here, ${\mathrm{d}\Omega}^2 \equiv {\mathrm{d}\theta}^2 + {{\sin ^2}\theta}\,{\mathrm{d}\varphi}^2$, $\left( {t,r,\theta,\varphi } \right)$ denotes a comoving coordinate system of the isotropic reference frame, and $a\!\left( t \right)$ is the scale factor, whose evolution is governed by a dynamical equation to be obtained later.
The exterior region ${{\cal M} ^+}$ is considered to be a static and spherically symmetric vacuum spacetime region, where the metric is currently expressed in the following generic form:
\begin{equation}
\label{EM} %"the Exterior Metric"
\mathrm{d}s_ + ^2 \equiv {g^{+}_{\mu \nu}} {\mathrm{d} x^{\mu}} {\mathrm{d} x^{\nu}} = - {f_1}\!\left( \tilde r \right) {\mathrm{d}{\tilde t}}^{\,2} + {{f_2}\!\left( \tilde r \right)}^{ - 1}{\mathrm{d}{\tilde r}}^2 + {{\tilde r}^2}{\mathrm{d}\Omega}^2\;.
\end{equation}
Here, $\left( {\tilde t,\tilde r,\theta,\varphi } \right)$ represents the Schwarzschild coordinate system, and ${f_1}\!\left( \tilde r \right)$ and ${f_2}\!\left( \tilde r \right)$ are undetermined metric functions.
It is important to note that $\theta$ and $\varphi$ are the common coordinates of the two regions, while $t$ and $r$ are particular to the interior region ${{\cal M} ^-}$, and $\tilde t$ and $\tilde r$ are specific to the exterior region ${{\cal M} ^+}$.

The interface $\varSigma$ in terms of interior coordinates is located at $r = {r_0}$, where ${r_0}$ is constant.
To seamlessly connect these two spacetime regions at the junction surface $\varSigma$, making it a continuous surface across the entire spacetime ${\cal M}$, specific junction conditions must be met.
In classical GR, the DI junction conditions dictate that the first fundamental form, namely the induced metric, and the second fundamental form, namely the extrinsic curvature, on either side of the boundary surface $\varSigma$ must be equal \cite{Darmois1927, Israel1966, Poisson2004}.
We will later assume that the aforementioned DI junction conditions remain valid for (quantum) modified theories, an assumption widely accepted in the literature \cite{Bojowald:2005qw, Piechocki:2020bfo, BenAchour:2020bdt, BenAchour:2020gon, Munch:2020czs, Schmitz:2020vdr, Lewandowski:2022zce, Yang:2022btw, Bobula:2023kbo, Bonanno:2023rzk, Giesel:2023hys, Cafaro:2024vrw, Bobula:2024jlh, Harada:2025cwd}.
Matching the interior region ${{\cal M} ^-}$ with the exterior region ${{\cal M} ^+}$ along the surface $\varSigma$ by applying the DI junction conditions gives rise to the following two relations between these two regions (see Appendix~\ref{AppendixA} for details):
\begin{gather}
a\!\left( t \right) {r_0} = {{\tilde r}\!\left ( t \right )}\;,
\label{Relation1} %"Relation 1"
\\
{f_1}\!\left( \tilde r \right) = {f_2}\!\left( \tilde r \right) = 1 - H^2 {\tilde r}^2 \;,
\label{Relation2} %"Relation 2"
\end{gather}
where $ H\!\left ( t \right ) := {\dot a\!\left ( t \right )}/{a\!\left ( t \right )} $ is the Hubble parameter, and the overdot denotes the derivative with respect to $t$.
Equation~\eqref{Relation1} provides a relation between the interior and exterior coordinates on the interface $\varSigma$, while Eq.~\eqref{Relation2} establishes a relation between the spacetime geometries of the two regions.

Equation~\eqref{Relation2} suggests that, within the current OS collapse framework, the metric of the exterior static and spherically symmetric vacuum spacetime must take the form of Eq.~\eqref{EM} with ${f_1}\!\left( \tilde r \right) = {f_2}\!\left( \tilde r \right)$.
The Schwarzschild metric in GR and the metrics of most regular BH models beyond GR satisfy this form.
On the other hand, some vacuum static spherically symmetric models that do not admit such a metric form (i.e., those whose metrics cannot be brought into the form of Eq.~\eqref{EM} with ${f_1}\!\left( \tilde r \right) = {f_2}\!\left( \tilde r \right)$ via coordinate transformations) are incompatible with this OS collapse scenario.
We henceforth denote ${f_1}\!\left( \tilde r \right) = {f_2}\!\left( \tilde r \right)$ by $f\!\left( \tilde r \right)$, whereby Eq.~\eqref{Relation2} can be rewritten as:
\begin{equation}
\label{f} %"the expression of f"
f\!\left( \tilde r \right) = 1 - H^2 {\tilde r}^2 \;,
\end{equation}
or
\begin{equation}
\label{H2} %"the expression of H^2"
H^2 = \frac{1 - f\!\left( \tilde r \right)}{{\tilde r}^2} \;.
\end{equation}
Note that the Hubble parameter $ H\!\left ( t \right ) $ of the interior dust region ${{\cal M}^ - }$ is a function of the interior time coordinate $t$.
However, on the right side of Eq.~\eqref{f}, it can also be viewed as a function of the exterior radial coordinate $\tilde r$ via the coordinate relation \eqref{Relation1}.
Conversely, the exterior coordinate $\tilde r$ on the right side of Eq.~\eqref{H2} can also be regarded as a function of the interior coordinate $t$ through Eq.~\eqref{Relation1}.
It is worth mentioning that Eqs.~\eqref{f} and \eqref{H2} play a significant role in linking the interior dynamics and the exterior solution of the collapsing dust ball, and more broadly, in bridging cosmological and BH geometries.

%%%%%%%%%%%%%% GR case %%%%%%%%%%%%%%
One can straightforwardly verify that, within the framework of classical GR, substituting the dynamical equation for the scale factor $a\!\left( t \right)$ in the standard FRW cosmology, which is named the Friedmann equation and reads:
\begin{equation}
\label{FE-GR} %"the Friedmann Equation in GR"
{H_{\text{GR}}}^{2}=\frac{8 \pi}{3} \rho \;,
\end{equation}
together with the metric function of the Schwarzschild spacetime:
\begin{equation}
\label{BH-GR} %"the metric function of BH in GR"
f\!\left( \tilde r \right) = {f_{\text{GR}}}\!\left( \tilde r \right) \equiv 1 - \frac{2M}{\tilde r} \;,
\end{equation}
into either Eq.~\eqref{f} or Eq.~\eqref{H2} leads to a self-consistent result.
Here the energy density $\rho$ for the dust-only model satisfies $\rho \propto {a\!\left ( t \right )}^{-3}$, and specifically in the current collapsing dust ball model, we have:
\begin{equation}
\label{ED} %"the expression of the Energy Density"
\rho\!\left ( t \right ) = \frac {M}{\frac{4}{3}\pi{{r_0}^3}{a\!\left ( t \right )}^3} \;.
\end{equation}
In the above equations, the constant $M$ denotes the mass of the dust ball region ${{\cal M}^ - }$ with the proper radius $a\!\left( t \right) {r_0} \equiv {\tilde r}\!\left( t \right)$.

%----------------------------------------------------
\subsection{Regular black hole with a Minkowski core}
\label{1.2} %"1.2 of the main body"
%----------------------------------------------------
%%%%%%%%%%%%%% Minkowski-core case %%%%%%%%%%%%%%
Before proceeding to the modified OS scenario in the next subsection, we first provide an overview of the Minkowski-core regular BH model, which will serve as the exterior geometry of that scenario.

A regular BH with an asymptotically Minkowski core was originally constructed at the phenomenological level by the authors of Ref.~\cite{LiXiang:2013sza}, designed to remove the curvature singularity and ensure the existence of a natural lower bound on the BH mass, whose metric function takes the form $f\!\left( \tilde r \right) = 1 - \left( 2M/{\tilde r} \right)\exp\!\left( -\alpha/{\tilde r}^2 \right)$.
It was shown that in this model the minimal BH mass is naturally determined by the existence of the horizon, and the Hawking temperature, heat capacity, and luminosity all vanish when the BH mass reaches this lower bound.
This offers a rather realistic and reasonable picture of the late stage of BH evaporation, which naturally terminates in a stable BH remnant.
Subsequently, this class of Minkowski-core regular BHs---characterized by the introduction of an exponential decay factor in the metric function, namely $f\!\left( \tilde r \right) = 1 - \left( 2M/{\tilde r} \right)\exp\!\left( -\epsilon\!\left( \tilde r \right) \right)$ with $\epsilon\!\left( \tilde r \right)>0$---attracted considerable interest and has since been generalized in a variety of forms \cite{Culetu:2013fsa, Culetu:2014lca, Balart:2014cga, Rodrigues:2015ayd, Simpson:2019mud, Ghosh:2014pba, Ghosh:2018bxg, Ling:2021olm}.
In these models, near the center, the exponential decay factor effectively suppresses the growth of the classical gravitational potential $M/{\tilde r}$, thereby leading to $f\!\left( \tilde r \right) \to 1$.
Furthermore, as ${\tilde r} \to 0$, the Riemann curvature tensor ${R_{\mu\nu\sigma}}^\rho \to 0$, resulting in the Einstein tensor $G_{\mu\nu} \to 0$, the Ricci scalar $R_{\text{ext}} \to 0$, and the Kretschmann scalar $K_{\text{ext}} \to 0$.
That is, this class of models exhibits a non-singular core that asymptotically approaches a Minkowski geometry.

Among the generalizations, the most recent one, proposed in Ref.~\cite{Ling:2021olm} and aimed at addressing the issue that the maximum value of the curvature scalar otherwise increases unboundedly with the BH mass, reads:
\begin{equation}
\label{BH-MC} %"the metric function of BH in the Minkowski-Core case"
f\!\left( \tilde r \right) = {f_{{\text{MC}}}}\!\left( \tilde r \right) \equiv 1 - \frac{{2M}}{\tilde r}{{\mathrm{e}^{-{\alpha _0}{M^x}/{{\tilde r}^n}}}} \;,
\end{equation}
with $\alpha_0 > 0$, $n>x\geq0$, and $n\geq1$.
The subscript “MC” represents “Minkowski-core”, distinguishing this model from the classical Schwarzschild BH and other modified BH models.
Here, the parameter $\alpha_0$ indicates the degree of deviation from the classical Schwarzschild BH; in other words, in the limit $\alpha_0 \to 0$, ${f_{{\text{MC}}}}\!\left( \tilde r \right)$ reduces to the classical case.
As for the remaining two parameters $n$ and $x$, Ref.~\cite{Ling:2021olm} mainly draws the following two conclusions.
Firstly, the maximum value of the curvature scalar is bounded above with increasing BH mass in the case of $x \ge n/3$ (especially the maximum curvature becomes mass independent when the equality $x = n/3$ holds), so that once $\alpha_0$ is appropriately fixed, the regular BH maintains a sub-Planckian curvature irrespective of its mass.
Secondly, for different choices of $n$ and $x$, this model can reproduce certain dS-core regular BHs at large scales, such as $(n,x)=(2,2/3)$ corresponding to the Bardeen BH and $(n,x)=(3,1)$ to the Hayward BH.
In particular, for the case $x = n/3 = 1$, ${f_{{\text{MC}}}}\!\left( \tilde r \right)$ becomes:
\begin{equation}
\label{BH-MC-1} %"the metric function of BH in the Minkowski-Core case (1)"
{f_{{\text{MC}}}}\!\left( \tilde r \right) \equiv 1 - \frac{{2M}}{\tilde r}{{\mathrm{e}^{-{\alpha _0}{M}/{{\tilde r}^3}}}} \;,
\end{equation}
whose asymptotic expansion at large scales ${{\tilde r} \gg \sqrt[3]{\alpha_0 M}}$ reads:
\begin{equation}
\label{AB-BH-MC-1} %"the Asymptotic Behavior of the metric function of BH in the Minkowski-Core case (1)"
{f_{{\text{MC}}}}\!\left( \tilde r \right) = 1 - \frac{{2M}}{\tilde r} \left ( 1 - \frac{{{\alpha_0}M}}{{\tilde r}^3} + {\cal O}\!\left( {\left( \frac{{{\alpha_0}M}}{{\tilde r}^3} \right)\!}^2 \right )\ \right )\;.
\end{equation}
It shares the same asymptotic behavior with the Hayward BH as well as the LQG BH (refer to Appendices~\ref{AppendixB} and \ref{AppendixC}) up to second-order terms in ${{{\alpha_0}M}}/{{\tilde r}^3}$ for a specific choice of $\alpha_0$\,\footnote{As detailed in Appendices~\ref{AppendixB} and \ref{AppendixC}, this requires the identification $\alpha_0 = 2l^2 = \alpha/2 \equiv 8\sqrt{3}\pi\gamma^3{l_{\text{P}}}^2 \approx 0.5832{l_{\text{P}}}^2$, where $l_{\text{P}} = 1$ denotes the Planck length.}.
This interesting identity of the large-scale asymptotic behavior of these three BH models was recently proposed in Ref.~\cite{Zhang:2024nny}.
In the following, the term “Minkowski-core BH” refers specifically to the model defined by Eq.~\eqref{BH-MC}, and in particular to its more specific version in Eq.~\eqref{BH-MC-1}.
Indeed, this BH model has since attracted considerable attention and become a significant platform for testing quantum gravity effects through astrophysical observations.
Extensive research has probed its observational signatures across various phenomena: BH shadows and accretion disk optics \cite{Ling:2022vrv, Zeng:2022yrm, Zeng:2023fqy, Xiong:2025hjn}, gravitational lensing \cite{Sodejana:2024mus, Guo:2024svn}, dynamical constraints from quasiperiodic oscillations and precession \cite{Guo:2025zca, Wu:2025xtn}, tidal Love numbers \cite{Wang:2025oek}, and perturbation analysis covering quasinormal modes and grey-body factors \cite{Zhang:2024nny, Tang:2024txx, Tang:2025mkk}, alongside gravitational waves from periodic orbits \cite{Gong:2025mne, Zare:2025aek} and late-time echoes \cite{Zhang:2025ygb}.
Such work collectively demonstrates that this BH model offers a tangible way to test quantum gravity effects against astronomical data.

%----------------------------------------------------
\subsection{New cosmology matching a Minkowski-core regular black hole}
\label{1.3} %"1.3 of the main body"
%----------------------------------------------------
In what follows, we consider a modified OS scenario, in which the exterior region ${{\cal M} ^+}$ is described by the Minkowski-core regular BH model given in Eq.~\eqref{BH-MC}.
Following the exterior-to-interior approach and substituting Eq.~\eqref{BH-MC} into Eq.~\eqref{H2} gives rise to the modified Friedmann equation for the corresponding interior cosmology:
\begin{equation}
\label{FE-MC} %"the Friedmann Equation in the Minkowski-Core case"
{H_{{\text{MC}}}}^2 = \frac{{8\pi\rho }}{3}{\exp\!\left( - {\alpha _0}{{\left( {\frac{{4\pi}}{3}\rho } \right)}^{\frac{n}{3}}}{M^{x - \frac{n}{3}}}\right)}\;.
\end{equation}
Here, the resulting cosmology accordingly adopts the subscript “MC” in its formulation.
It is easy to see that ${H_{{\text{MC}}}}^2$ recovers its classical counterpart ${H_{{\text{GR}}}}^2$ in the classical limit where $\alpha_0 \to 0$.
Equation~\eqref{FE-MC} also indicates that, in addition to the energy density $\rho$ and the three model parameters $\alpha_0 $, $x$, and $n$ inherited from the Minkowski-core BH, the Hubble parameter ${H_{{\text{MC}}}}$ is also related to the constant $M$.
However, in the context of cosmology, the constants $r_0$ and $M$ in Eq.~\eqref{ED} should be interpreted as the comoving radius of an arbitrary spherical region in the universe and the total mass enclosed within it, respectively.
Due to the spatial homogeneity of the universe, the choice of such a spherical region is arbitrary, and its total mass $M$ changes with its comoving radius $r_0$.
Therefore, the evolution of the scale factor ${a\!\left ( t \right )}$ should be independent of $r_0$ and $M$, implying that these two constants are not expected to appear in the evolution equation of the scale factor ${a\!\left ( t \right )}$.
Given these arguments, it is reasonable to require the parameters $x$ and $n$ to satisfy the condition $n = 3x$ such that $M$ drops out of the modified Friedmann equation \eqref{FE-MC}.
Remarkably, this requirement coincides with the condition in the BH context that the maximum curvature scalars are mass independent.
Next, we proceed to consider the specific case $n = 3x = 3$, in which Eq.~\eqref{FE-MC} reduces to:
\begin{equation}
\label{FE-MC-1} %"the Friedmann Equation in the Minkowski-Core case (1)"
{H_{{\text{MC}}}}^2 = \frac{{8\pi\rho }}{3}{\mathrm{e}^{- {\frac{{4\pi{\alpha _0}}}{3}\rho }}}\;,
\end{equation}
which can be expanded in the large-scale, low-density regime as follows:
\begin{equation}
\label{AB-FE-MC-1} %"the Asymptotic Behavior of the Friedmann Equation in the Minkowski-Core case (1)"
{H_{{\text{MC}}}}^2 = \frac{8\pi}{3}\rho \left ( 1 - \frac{4\pi{\alpha _0}}{3}\rho + {\cal O}\!\left( {\left( \frac{4\pi{\alpha _0}}{3}\rho \right)\!}^2 \right )\ \right )\;.
\end{equation}
As expected, it coincides with that of the Hayward cosmology and LQC (refer to Appendices~\ref{AppendixB} and \ref{AppendixC}) up to terms of ${\cal O}\!\left( {\left( {4\pi{\alpha _0}\rho}/{3} \right)}^2 \right )$ upon appropriately setting $\alpha_0$.
In the next section, we shall focus on the newly obtained cosmological model with $n = 3x = 3$, studying its dynamical features and carrying out a series of comparative analyses with LQC, the Hayward cosmology, as well as the classical standard cosmology.

%====================================================
\section{Dynamics of the non-singular universes}
\label{2} %"the 2nd section of the main body"
%----------------------------------------------------
In the previous section, we brought up the modified Friedmann equation for the new cosmological model matched to the Minkowski-core regular BH model by means of the OS dust collapse framework.
Besides, the modified Friedmann equations for the other two non-singular cosmological models---LQC and the Hayward cosmology---have been provided in Appendices~\ref{AppendixB} and \ref{AppendixC}, respectively, and the standard Friedmann equation of classical GR is in subsection~\ref{1.1}.
By making use of these equations, in this section, we investigate and compare the dynamical evolution of these models in two distinct settings: one is the dust-only universe scenario (subsection~\ref{2.1}) naturally arising from the OS framework, and the other involves a universe filled with a massive scalar field (subsection~\ref{2.2}).
For the new model specifically, the analysis is then extended to other cosmological fluids and a multi-fluid description incorporating a positive cosmological constant (subsection~\ref{2.3}), and the robustness of the asymptotic past is examined within a dynamical systems framework (subsection~\ref{2.4}).

In order to facilitate a unified and concise treatment in the subsequent analysis, the Friedmann equations for various cosmological models can be uniformly written in a general form:
\begin{equation}
\label{FE-U} %"the Friedmann Equation in a Unified form"
{H}^2 = F\!\left ( \rho \right )\;.
\end{equation}
Here, the function $F\!\left ( \rho \right )$ in the aforementioned four models takes the following forms:
\begin{subequations}
\label{F-rho} %"the function F(rho)"
\begin{align}
{F_{\text{GR}}}\!\left( \rho  \right) &= \frac{{8\pi}}{3}\rho\;, \label{F-rho-a}\\
{F_{{\text{LQG}}}}\!\left( \rho  \right) &= \frac{{8\pi}}{3}\rho \left( {1 - \frac{\rho }{{{\rho _{\text{c}}}}}} \right)\;, \label{F-rho-b}\\
{F_{{\text{Hayward}}}}\!\left( \rho  \right) &= \frac{{8\pi\rho }}{{3 + 8\pi{l^2}\rho }}\;, \label{F-rho-c}\\
{F_{{\text{MC}}}}\!\left( \rho  \right) &= \frac{{8\pi\rho }}{3}{{\mathrm{e}} ^{ - \frac{{4\pi{\alpha_0}}}{3}\rho }}\;. \label{F-rho-d}
\end{align}
\end{subequations}
In Eqs.~\eqref{F-rho-b} and \eqref{F-rho-c}, $\rho _{\text{c}}$ and $l$ are constants, with further details provided in Appendices~\ref{AppendixB} and \ref{AppendixC}.

%----------------------------------------------------
\subsection{Dust-only scenario}
\label{2.1} %"2.1 of the main body"
%----------------------------------------------------
In this subsection, we analyze the scenario of a dust-only universe, whose contents consist solely of dust and contain no other kinds of matter.
In fact, the spacetime geometry of the interior region ${{\cal M} ^-}$ of the OS collapsing dust ball model studied in the previous section can be regarded as the time-reversed version of an expanding dust-only universe.
For the dust-only universe, as mentioned in subsection~\ref{1.1}, the energy density $\rho$ is inversely proportional to the cube of the scale factor $a\!\left ( t \right )$, which can still be expressed as Eq.~\eqref{ED}, where the meanings of the constants $r_0$ and $M$ have already been clarified in subsection~\ref{1.3}.

%----------------------------------------------------
\subsubsection{Dynamical evolution}
\label{2.1.1} %"2.1.1 of the main body"
\begin{figure*}[ht]
\centering
\vspace{1em}
\includegraphics[width=0.98\textwidth]{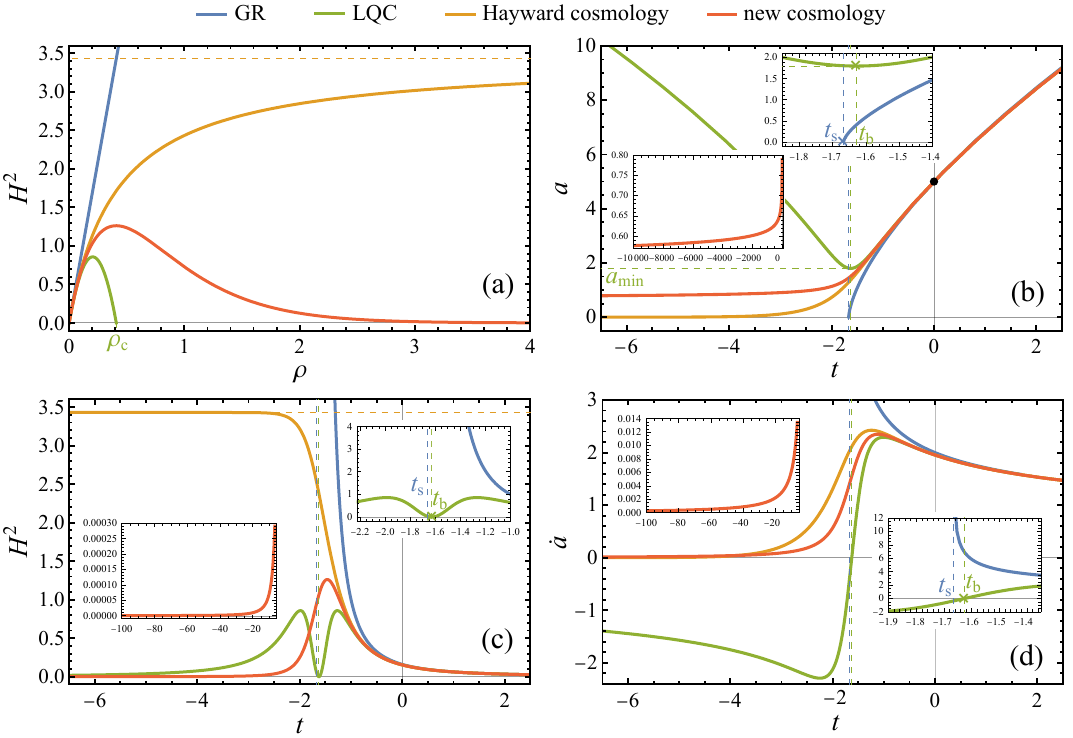}
\caption{Plots of (a) $H\!\left ( \rho \right )^2$, (b) $a\!\left ( t \right )$, (c) $H\!\left ( t \right )^2$, and (d) $\dot a\!\left ( t \right )$ for the four cosmologies. Here, $r_0 = 1$, $M = 10$, and $a_0 = 5$.}
\label{H&a}
\end{figure*}

%----------------------------------------------------
Combining the Friedmann equation \eqref{FE-U} with Eq.~\eqref{ED} yields:
\begin{equation}
\label{ODE1-1} %"the 1st Ordinary Differential Equation 1"
{H}^2 \equiv \left ( \frac{\dot a}{a} \right )^2  = F\!\left ( \rho \!\left ( a \right ) \right )\;,
\end{equation}
or
\begin{equation}
\label{ODE1-2} %"the 2nd Ordinary Differential Equation 1"
\frac{\mathrm{d} a}{\mathrm{d} t}  = \pm \, a \sqrt{F\!\left ( \rho \!\left ( a \right ) \right )}\;.
\end{equation}
In Eq.~\eqref{ODE1-2}, the sign “$+$” designates the expansion scenario, whereas the sign “$-$” indicates the contraction one.
Obviously, the obtained equation is an autonomous (i.e., with no explicit dependence on the independent variable $t$) first-order ordinary differential equation (ODE) for the scale factor $a\!\left ( t \right )$.
Rearranging Eq.~\eqref{ODE1-2} leads to:
\begin{equation}
\label{ODE1-3} %"the 3rd Ordinary Differential Equation 1"
{\mathrm{d} t} = \frac{\mathrm{d} a}{{\pm \, a \sqrt{F\!\left ( \rho \!\left ( a \right ) \right )} }}\;.
\end{equation}
Given an initial condition $a\!\left ( 0 \right ) = a_0$, we can integrate both sides to obtain:
\begin{equation}
\label{t-a} %"t(a)"
t\!\left ( a \right ) = \int_{a_0}^{a} \frac{\mathrm{d} a_1}{\pm \, a_1 \sqrt{F\!\left ( \rho \!\left ( a_1 \right ) \right )}}\;.
\end{equation}
In principle, one can obtain $a\!\left( t \right)$ by inverting the function $t\!\left( a \right)$.
For the GR and LQC cases, the resulting equations remain simple enough to allow for an explicit analytical solution for $a\!\left(t\right)$; by contrast, the Hayward case, and our new model even more so, give rise to much more complex integrals, necessitating numerical techniques to obtain $a\!\left(t\right)$, and then $\dot a\!\left( t \right)$ as well as ${H\!\left( t \right)}^2$.
The obtained results are depicted in Fig.~\ref{H&a}(b)--(d).
For reference, the curves of $H\!\left (\rho\right )^2$ provided in Eq.~\eqref{FE-U} with Eq.~\eqref{F-rho} are also plotted in Fig.~\ref{H&a}(a).

In Fig.~\ref{H&a}(b), the black dot marks the common initial condition $a\!\left( 0 \right) = a_0 = 5$ shared by all four cosmological models, and the expanding phase in accordance with the present universe is chosen at the initial time.
Across all panels of Fig.~\ref{H&a}, colored thin dashed lines represent a variety of reference lines associated with the curves (shown as thick solid lines) of the same color and hence with the corresponding models.
Colored cross markers and colored symbols also follow the color coding of the respective models.
In particular, in panels (b)--(d), inset panels with adjusted plotting ranges are provided to more clearly display the meanings of the blue symbol $t_\text{s}$ and the green symbol $t_\text{b}$---the former stands for the time of the big bang singularity in GR satisfying $a\!\left( t_\text{s} \right) = 0$, while the latter indicates the time of the big bounce in LQC satisfying $a\!\left( t_\text{b} \right) = a_\text{min}$.
Furthermore, for the new cosmology in panels (b)--(d), the portions of the curves that extend beyond the left edges of the main panels and cover much broader ranges along the negative time axis are shown as inset panels within each respective panel, in order to better illustrate the features of the earlier-time evolution in the new cosmology.

%【分析图。】
% 晚期阶段各模型相互重合
As is apparent from Fig.~\ref{H&a}, in agreement with the analysis in subsection~\ref{1.3} (along with Appendices~\ref{AppendixB} and \ref{AppendixC}), the curves of the three non-singular cosmologies almost coincide with that of classical GR in the late-time (low-density) regime; even in the slightly earlier regime, where the three non-singular cosmologies start to deviate marginally from GR, their curves still nearly overlap with each other, since these three cosmologies share the same asymptotic behavior up to higher-order corrections at large scales; in contrast, in the early-time (high-density) regime, these three non-singular cosmologies exhibit behaviors that differ drastically not only from the classical case but also from one another.

% 早期阶段各模型各不相同
% GR
Focusing first on the GR case, one observes that when evolving backward toward the finite big-bang-singularity time $t_\text{s} = -5/3 \approx -1.667$, the scale factor $a\!\left( t \right)$ rapidly decreases to zero, leading to a divergent energy density $\rho$, while both the rate of change of the scale factor $\dot a$ and the Hubble parameter $H$ blow up.
% LQC
Turning to the case of LQC, when evolving backward to the finite big-bounce time $t_\text{b} \approx -1.627$, $\dot a$ and $H$ decrease to zero (thereafter changing sign and becoming negative), the scale factor $a\!\left( t \right)$ reaches its minimum value $a_\text{min} \approx 1.800$, and the energy density $\rho$ attains the critical density ${\rho _{\text{c}}} \approx 0.4094$, at which point the evolution smoothly transitions to the time-reversed counterpart.
% the Hayward & Minkowski-core models
In contrast to the GR case, where the energy density $\rho$ diverges at a finite past time $t_\text{s}$, and to LQC, where $\rho$ is bounded above by a critical density $\rho_{\text{c}}$, both the Hayward and the new cosmologies feature a divergence of $\rho$ in the infinite past.
In the early-time (high-density) regime, despite roughly sharing the above feature, these two cosmologies exhibit markedly different behaviors in detail: in the Hayward cosmology, both $a\!\left(t\right)$ and $\dot a$ rapidly decrease toward zero (with $\rho$ increasing toward infinity due to the former), driving $H^2$ to quickly approach the constant $1/l^2 \approx 3.430$; whereas in the new cosmology, although $\dot a$ also rapidly decreases toward zero, $a\!\left( t \right)$ instead undergoes a very slow but persistent decrease, with the consequence that $H^2$ quickly approaches zero.

% 其他特征，按图区分
From Fig.~\ref{H&a}(d), one observes that, as mentioned in the previous paragraph, in LQC the rate of change of the scale factor $\dot a$ vanishes at a finite time and then changes sign, thereby giving rise to the so-called big bounce.
In contrast, in the Hayward and the new cosmologies, $\dot a$ never reaches zero within finite time, and thus no such bounce occurs.
Additionally, as can be seen from panels (b)--(d), in sharp contrast to the classical GR case, in the three non-singular cosmologies the affine parameter $t$ along timelike geodesics spans the full range $(-\infty, \infty)$ throughout the entire evolution, indicating that these cosmological spacetimes are timelike geodesically complete.

%----------------------------------------------------
\subsubsection{Curvature scalars}
\label{2.1.2} %"2.1.2 of the main body"
\begin{figure*}[ht]
\centering
\vspace{1em}
\includegraphics[width=0.98\textwidth]{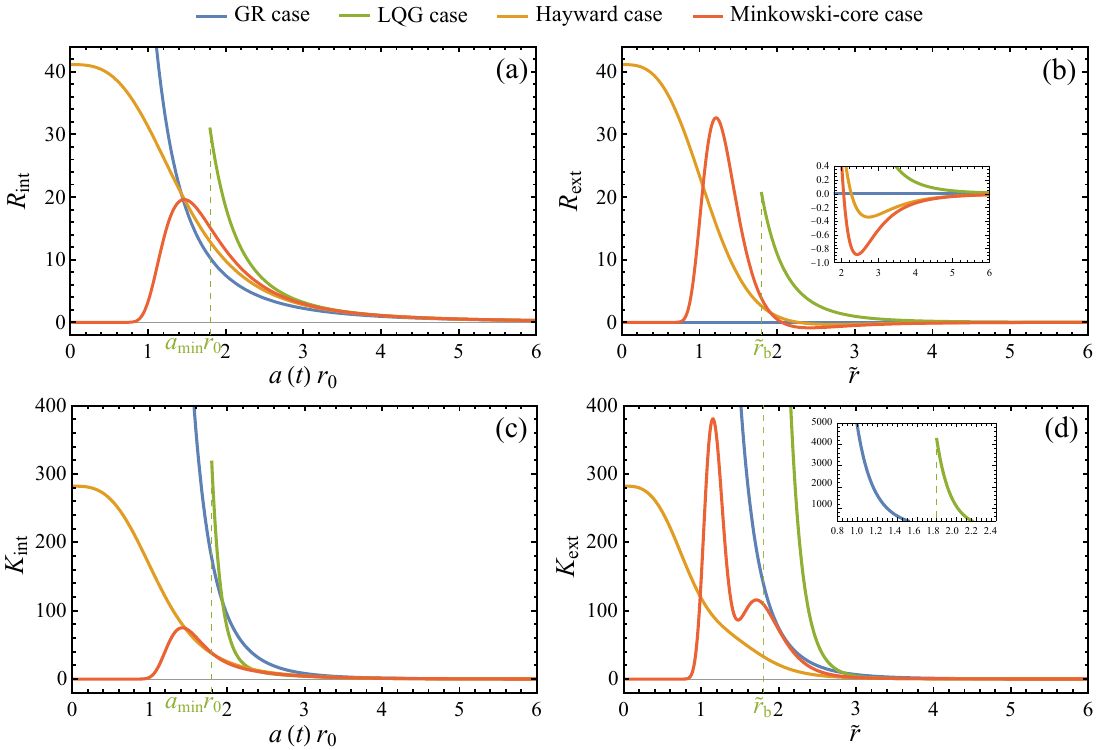}
\caption{(a) and (c): $R_{\text{int}}$ and $K_{\text{int}}$ (for the four cosmological models) versus $a\!\left( t \right) {r_0}$, respectively; (b) and (d): $R_{\text{ext}}$ and $K_{\text{ext}}$ (for the four corresponding BH models) versus ${\tilde r}$, respectively. Here, $r_0 = 1$ and $M = 10$.}
\label{R&K} 
\end{figure*}

%----------------------------------------------------
% 给出曲率标量图
To further inquire into how the three non-singular cosmological models resolve the spacetime singularity, we compute the Ricci and Kretschmann scalars for these models and the classical standard cosmology varying with respect to the proper radius, as well as those for their corresponding BH models varying with respect to the radial coordinate, which can be calculated by the following formulas:
\begin{align}
\label{R_int}
{R_{\text{int}}} &= \frac{6\left({{\dot a}^2} + a\ddot a \right)}{a^2} = \frac{6\left( t'\!\left( a \right) - at''\!\left( a \right) \right)}{{a^2}{t'\!\left( a \right)}^3}\;, \\
\label{K_int}
{K_{\text{int}}} &= \frac{12\left( {{\dot a}^4} + {a^2}{{\ddot a}^2} \right)}{a^4} = \frac{12\left( {a^2}{t''\!\left( a \right)}^2 + {t'\!\left( a \right)}^2 \right)}{{a^4}{t'\!\left( a \right)}^6}\;, \\
\label{R_ext}
{R_{\text{ext}}} &= \frac{2 - 2f\!\left( {\tilde r} \right) - 4{\tilde r}f'\!\left( {\tilde r} \right) - {{\tilde r}^2}f''\!\left( {\tilde r} \right)}{{\tilde r}^2}\;, \\
\label{K_ext}
{K_{\text{ext}}} &= \frac{4 - 8f\!\left( {\tilde r} \right) + {4f\!\left( {\tilde r} \right)}^2 + {4{{\tilde r}^2}f'\!\left( {\tilde r} \right)}^2 + {{\tilde r}^4}{f''\!\left( {\tilde r} \right)}^2}{{\tilde r}^4}\;.
\end{align}
Here, for convenience, we follow the setup of the OS scenario in the previous section (thereby using the subscripts “int” and “ext” to denote (interior) cosmological spacetimes and (exterior) BH spacetimes, respectively) together with the corresponding coordinate notation.
In addition, the evaluation of ${R_{\text{int}}}$ and ${K_{\text{int}}}$ requires further substituting Eqs.~\eqref{ODE1-3} and \eqref{ED}.
The resulting plots are shown in Fig.~\ref{R&K}.

%【分析图。】
% GR
From Fig.~\ref{R&K}, one can see that within the framework of GR, apart from the fact that the Ricci scalar ${R_{\text{ext}}}$ of the Schwarzschild BH spacetime vanishes identically throughout the whole space, the Kretschmann scalar ${K_{\text{ext}}}$ of the BH spacetime, as well as ${R_{\text{int}}}$ and ${K_{\text{int}}}$ of the standard cosmological spacetime, all grow steeply and tend to diverge as the scale approaches zero.
% LQG
In contrast, within the framework of LQG, both the Ricci and Kretschmann scalars of both the cosmological and BH spacetimes increase rapidly near the Planck scale; however, due to the existence of the minimal scales $a_{\text{min}}r_0$ and ${\tilde r}_{\text{b}}$---where the bounce occurs---the curvature scalars stay bounded and hence avoid divergence.
% Hayward & Minkowski-core cases
As for the Hayward and the Minkowski-core cases, one observes that as the scale decreases toward zero, in terms of the overall trend, in the former case the two curvature scalars of both the cosmological and BH spacetimes first increase significantly, then level off, and eventually reach a finite value, consistent with the description in Appendix~\ref{AppendixC}; whereas those in the latter case first increase markedly, then rapidly decrease to nearly zero, and finally vanish, as described in subsections~\ref{1.2} and \ref{1.3}.
There are two particular details.
First, in panel (b), the Ricci scalar ${R_{\text{ext}}}$ of both the Hayward BH and the Minkowski-core BH spacetimes is negative beyond a certain small scale, decreasing from $0^{-}$ at infinity and then turning upward to cross zero, and the magnitude of its minimum is much smaller than the maximum subsequently reached at smaller scales.
Second, in panel (d), the Kretschmann scalar ${K_{\text{ext}}}$ of the Minkowski-core BH spacetime exhibits two peaks, one higher and one lower.
However, further investigations of these two features lie beyond the scope of this work.

In summary, whereas the classical big-bang singularity is replaced by a big bounce in LQC, and by an asymptotically dS epoch in the Hayward cosmology, it is replaced by an asymptotically Minkowski epoch in our new cosmology.
These very-early-time behaviors of the cosmological models agree well with the behaviors of their respective BH counterparts in the small-scale regime near the center.

%----------------------------------------------------
\subsection{Scalar-field scenario}
\label{2.2} %"2.2 of the main body"
%----------------------------------------------------
In this subsection, we turn to another scenario, where the universe is filled with a massive scalar field $\phi$ with mass $m$ and a potential $V\!\left( \phi  \right) = \frac{1}{2}{m^2}{\phi ^2}$.
The energy density $\rho$ and the pressure $p$ of the scalar field $\phi$ are given by:
\begin{subequations}
\label{rho&p} %"the expressions of rho & p"
\begin{align}
\rho &= \frac{1}{2}{{\dot \phi }^2} + V\!\left( \phi  \right)\;, \label{rho&p-a}\\
p &= \frac{1}{2}{{\dot \phi }^2} - V\!\left( \phi  \right)\;. \label{rho&p-b}
\end{align}
\end{subequations}
The conservation law is:
\begin{equation}
\label{CL} %"the Conservation Law"
\dot \rho  + 3H\left( {\rho  + p} \right) = 0\;.
\end{equation}
By inserting Eq.~\eqref{rho&p} into Eq.~\eqref{CL}, one arrives at the Klein--Gordon equation:
\begin{equation}
\label{KGE} %"the Klein--Gordon Equation"
\ddot \phi + 3H\dot \phi + V' = 0\;,
\end{equation}
where $V' \equiv {\mathrm{d}V}/{\mathrm{d}\phi}$.

\begin{figure*}[ht]
\centering
\vspace{1em}
\includegraphics[width=0.98\textwidth]{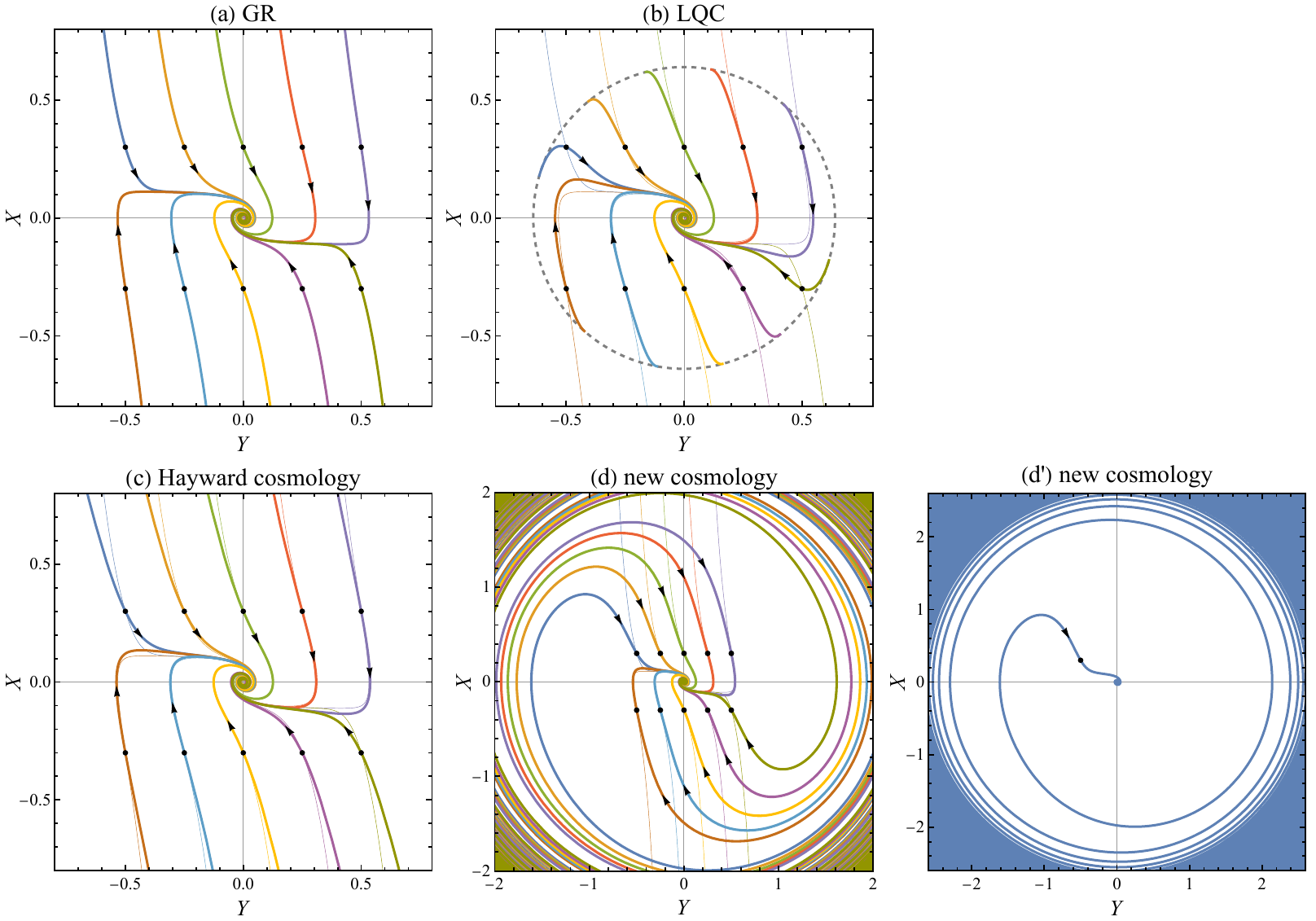}
\caption{Phase trajectories of the scalar field in the four cosmologies. Here, $m = 1$.}
\label{Dyn}
\end{figure*}

In this context, the time evolution of the scalar field $\phi$ is of greater concern.
To simplify the description, we introduce the following new phase-space variables:
\begin{subequations}
\label{NVs} %"New Variables"
\begin{align}
X &:= \frac{{\dot \phi }}{{\sqrt 2 }}\;, \label{NVs-a}\\
Y &:= \sqrt V \equiv \frac{{m\phi }}{{\sqrt 2 }}\;. \label{NVs-b}
\end{align}
\end{subequations}
Combining the above definitions with Eq.~\eqref{rho&p-a} immediately gives:
\begin{equation}
\label{rho-NVs} %"the relationship between rho and New Variables"
\rho  = {X^2} + {Y^2}\;.
\end{equation}
Then, substituting Eqs.~\eqref{FE-U}, \eqref{NVs}, and \eqref{rho-NVs}, together with the expression of the potential function $V\!\left( \phi  \right)$, into Eq.~\eqref{KGE} yields:
\begin{equation}
\label{ODE2} %"the Ordinary Differential Equation 2"
\ddot \phi =  \mp ~ 3\sqrt {F\!\left( {{X^2} + {Y^2}} \right)} \sqrt 2 X - \sqrt 2 mY\;,
\end{equation}
where the sign “$-$” corresponds to the expanding branch and the sign “$+$” to the contracting one.
Following this, by differentiating Eqs.~\eqref{NVs} with respect to $t$ and incorporating Eq.~\eqref{ODE2}, we obtain the system of dynamical equations in terms of the new variables as follows:
\begin{subequations}
\label{ODEs2} %"the Ordinary Differential Equations 2"
\begin{align}
{\dot X} &= \mp ~ 3\sqrt {F\!\left( {{X^2} + {Y^2}} \right)} X - mY\;, \label{ODEs2-a}\\
{\dot Y} &= mX\;. \label{ODEs2-b}
\end{align}
\end{subequations}
It is evident that this system also constitutes an autonomous system.
In effect, the second-order nonlinear ODE \eqref{KGE} for $\phi\!\left(t\right)$ has been reduced to a system of nonlinear first-order ODEs \eqref{ODEs2} for the new variables.
Subsequently, after substituting the specific forms \eqref{F-rho} of $F\!\left(\rho\right)$ in the four models into Eq.~\eqref{ODEs2}, we numerically solve the resulting systems and obtain the phase trajectories in the $Y$--$X$ plane, as displayed in Fig.~\ref{Dyn}.
Among them, the phase trajectories in GR and LQC were originally presented in Refs.~\cite{Belinsky:1985zd} and \cite{Singh:2006im}, respectively.
It is noteworthy that, according to Eq.~\eqref{rho-NVs}, the distance from a point in the $Y$--$X$ phase plane to the origin is precisely equal to $\sqrt{\rho}$.
This means that the scalar-field energy density $\rho$ corresponding to a given point on the phase trajectories can be directly ascertained from the distance of that point to the origin in the phase plane.

Notably, the trajectories for GR, shown in Fig.~\ref{Dyn}(a), are also plotted as thin solid lines in the other three panels (b)--(d), serving as a reference for direct visual comparison with the non-singular cosmologies.
Across all panels of Fig.~\ref{Dyn}, black dot marks indicate the initial conditions for numerical solutions of the corresponding trajectories, and phase trajectories starting from the same initial condition are displayed in the same color.
Black arrows on the trajectories identify the direction of time evolution; here, without loss of generality, we choose to display the expansion phase.
The gray thick dashed circle in panel (b) does not represent a phase trajectory, but rather the boundary beyond which the system reaches the critical energy density $\rho_{\text{c}}$ and undergoes a bounce in LQC.
In panel (d), the phase trajectories for the new cosmology appear somewhat cluttered; to facilitate visual clarity, a single representative trajectory is extracted and shown separately in panel (d$'$), illustrating the characteristic evolution of an individual solution in the new cosmology.

%【分析图。】
% 近晚期共同点——吸引子和暴涨相
It is apparent from Fig.~\ref{Dyn} that both the Hayward and the new cosmologies exhibit an attractor solution similar to that in GR and LQC (see, e.g., Refs.~\cite{Belinsky:1985zd, Felder:2002jk, mukhanov2005physical, Singh:2006im} for details).
Specifically, phase trajectories with different initial conditions are all drawn toward the inflationary phase (characterized by ${\mathrm{d} X}/{\mathrm{d} Y} \approx 0$) as time evolves and subsequently undergo numerous damped oscillations before eventually settling at the origin of the phase plane, which corresponds to a stable node.
This fact implies that, within the present scalar field setup, inflation is also a generic feature in the Hayward and the new cosmologies, just as the authors of Refs.~\cite{Belinsky:1985zd} and \cite{Singh:2006im} concluded for GR and LQC, respectively.
It follows that inflation can be harmoniously incorporated into all three non-singular cosmological frameworks beyond GR.
% 早期差异
Although the phase trajectories in the three non-singular cosmologies behave almost identically to those in GR at late times, these four cosmologies differ significantly from GR during the early stages of evolution.
The most distinctive feature of the LQC trajectories is the presence of the bounce boundary, as discussed in detail in Ref.~\cite{Singh:2006im}.
In the Hayward cosmology, the phase trajectories deviate increasingly from the GR trajectories as the evolution is traced further back in time.
Finally, in the new cosmology, starting from indefinitely early times up to a certain moment prior to the onset of inflation, the phase trajectories exhibit continuous damped oscillations---the earlier the time, the less the amplitude decays per oscillation cycle, manifesting in the phase plane as trajectories becoming denser farther away from the origin.
It is worth noting that during the late-time damped oscillations of the scalar field that eventually vanish, the phase trajectories become increasingly dense as they approach the origin.
This corresponds to a decrease in the energy density at a progressively slower rate, which implies the decelerating expansion of the universe.
One finds that the scalar field in the new cosmology exhibits an intriguing symmetry between its behavior in the asymptotic past and future.

%----------------------------------------------------
\subsection{Other cosmological fluids and multi-fluid scenario}
\label{2.3} %"2.3 of the main body"
%----------------------------------------------------
It should be noted, however, that the dynamical analyses of the four cosmological models carried out thus far in this section are based on a highly idealized cosmological setup.
As first discovered in high-redshift Type Ia supernova observations in 1998 \cite{SupernovaSearchTeam:1998fmf, SupernovaCosmologyProject:1998vns}, the present universe is undergoing accelerating rather than decelerating expansion, which indicates that our late-time universe is dominated by some form of dark energy with negative pressure.
In reality, in modern cosmology, the contents of the universe are typically deemed to include at least ordinary matter, radiation, dark matter, and dark energy, with different components dominating at different epochs.
By contrast, the scenarios considered in the preceding two subsections involve only dust matter or a scalar field.
While such idealizations are sufficient to capture the essential non-singular features of the models under consideration, they clearly fall short of faithfully capturing the full content of the real universe.

In what follows, we first consider other cosmological fluids.
By introducing the equation-of-state parameter $w := p/\rho$, the conservation law \eqref{CL} can be rewritten as:
\begin{equation}
\label{CL-2} %"the 2nd version of the Conservation Law"
\dot \rho  + 3H \rho \left( {1 + w} \right) = 0\;.
\end{equation}
Ordinary and dark matter both typically have $w \simeq 0$ and are commonly treated as dust ($w_{\text{d}}=0$).
For radiation, $w_{\text{r}}=1/3$.
As for dark energy, which dominates the current late-time universe, astronomical observations are consistent with $w \simeq -1$, with the simplest and most favored candidate being a cosmological constant $\varLambda$, for which $w_{\varLambda}=-1$.
In the following, we consider three representative extreme scenarios---containing only dust, radiation, or a cosmological constant---and integrate Eq.~\eqref{CL-2} to obtain the dependence of $\rho$ on the scale factor $a$, respectively as follows:
\begin{subequations}
\label{rho-a} %"the function rho(a)"
\begin{align}
\rho_{\text{d}} &\propto a^{-3}\;, \label{rho-a-dust}\\
\rho_{\text{r}} &\propto a^{-4}\;, \label{rho-a-radiation}\\
\rho_{\varLambda} &= \text{const}\;. \label{rho-a-Lambda}
\end{align}
\end{subequations}
Substituting these relations into the respective $F\!\left ( \rho \right )$ \eqref{F-rho} in the Friedmann equation \eqref{FE-U} yields the dynamical behavior of each cosmology in each limiting scenario.
The first relation \eqref{rho-a-dust} is precisely the one already used in subsections~\ref{1.1} and \ref{2.1}.
Within the framework of classical GR, since the square of the Hubble parameter is proportional to the energy density, it is evident that in the early universe, i.e., when the scale factor is small, radiation dominates, while in the late universe, i.e., when the scale factor is large, dark energy (the cosmological constant) dominates.
In modified models, however, this picture may no longer hold, as we will demonstrate later.
For the radiation-only scenario, a straightforward calculation and inspection shows that the distinctive features of the temporal evolution are almost identical to those in the dust-only case (as presented in subsection~\ref{2.1}) for each of the four models, and particularly each model retains its characteristic early-time behavior---the equation of state only affects the quantitative approach rate, not the qualitative nature of the early-time behavior.
As for the scenario involving only a cosmological constant, substituting Eq.~\eqref{rho-a-Lambda} into the Friedmann equation \eqref{FE-U} with \eqref{F-rho}, one finds that for all four models the Friedmann equation reduces to $H^2 = F\!\left ( \rho_{\varLambda} \right ) \equiv \text{const}$, giving rise to the same dS expansion and rendering the three modified models essentially indistinguishable from the classical GR case in such a scenario.
This indicates the modified models exhibit almost no difference from the GR case during the cosmological-constant-dominated era.

We now proceed to examine a more realistic scenario where multiple fluid components coexist within our new cosmological model.
In view of the fact that $\rho_{\varLambda}$ does not evolve with time, we distinguish it from the dynamical part contributed by other matter components, $\rho\!\left ( t \right ) = \rho_{\text{d}}\!\left ( t \right ) + \rho_{\text{r}}\!\left ( t \right )$, and treat it directly as an independent positive cosmological constant term in the Friedmann equation:
\begin{equation}
\label{FE-MC-Lambda} %""
{H_{\text{MC},\varLambda}}^2 = F_{\text{MC}}\!\left ( \rho\!\left ( t \right ) \right ) + \frac{\varLambda}{3} = \frac{8\pi\rho\!\left ( t \right )}{3}{\mathrm{e}} ^{ - \frac{{4\pi{\alpha_0}}}{3}\rho\left ( t \right ) } + \frac{\varLambda}{3}\;.
\end{equation}
This equation implies two distinct asymptotic regimes.
(i) Late-time, low-density regime ($8\pi\rho\!\left ( t \right ) \ll \varLambda$): the $\rho\!\left ( t \right )$ term becomes negligible and Eq.~\eqref{FE-MC-Lambda} approaches ${H_{\text{MC},\varLambda}}^2 \approx \varLambda/3$, corresponding to an eternally accelerating dS phase, in agreement with the conclusion drawn in the preceding paragraph.
(ii) Early-time, high-density regime ($8\pi\rho\!\left ( t \right ) \gg \varLambda$).
Although the exponential dynamical term dominates the right-hand side at intermediate densities, in the strict infinite-past limit $\rho\!\left ( t \right ) \to \infty$ the exponential suppression $F_{\text{MC}} \to 0$ kills the $\rho\!\left ( t \right )$ contribution, leaving ${H_{\text{MC},\varLambda}}^2 \to \varLambda/3$.
Thus, the universe approaches a dS phase---not Minkowski---in the infinite past, with ${H_{\text{MC},\varLambda}} \to \sqrt{\varLambda/3}$ and non-vanishing curvature scalars.
The defining exponential suppression of this model remains operative, but the presence of $\varLambda$ shifts the early-time asymptotics from Minkowski to dS.
In other words, the symmetry between the early- and late-time regimes of the pure new cosmology transitions from asymptotically Minkowski to asymptotically dS when a positive cosmological constant is present.

The same reasoning applies, \textit{mutatis mutandis}, to the Hayward cosmology and LQC: each model retains its characteristic early-universe behavior (early dS plateau for the Hayward cosmology, bounce for LQC) in the presence of multiple fluids, because the dominant energy density at early times still enters the respective $F\!\left ( \rho \right )$ and drives the defining non-singular feature.
It is instructive to contrast the new cosmology with the Hayward cosmology in this regard.
The Hayward BH possesses an \textit{intrinsically} asymptotically dS core, characterized by an effective cosmological constant $\varLambda_{\text{eff}} = 3/l^2$ near $\tilde{r} \to 0$ (see Appendix~\ref{AppendixC}), so that its matching cosmology displays an asymptotically dS epoch at very early times even without an external $\varLambda$.
In the Minkowski-core BH, which is matched with the new cosmology, by contrast, the non-singular core is Minkowskian; consequently, a pure new cosmology ($\varLambda=0$) approaches Minkowski spacetime in both the asymptotic past and future.
When a positive $\varLambda$ is introduced to account for the observed late-time cosmic acceleration, the exponential suppression term at high densities leaves ${H_{\text{MC},\varLambda}}^2 \to \varLambda/3$ in the infinite past---shifting the early-time asymptotics from Minkowski to dS, as discussed above.
The incorporation of $\varLambda$ is required in both models to match late-time observations, and in neither case does it reintroduce a primordial singularity: both cosmologies remain globally non-singular.
With the same extrinsically supplied $\varLambda$, the two cosmologies differ in the degree of their early-time dS behavior: the early-time limit of the new cosmology, ${H_{\text{MC},\varLambda}}^2 \to \varLambda/3$, simply coincides with its late-time one, in that the ``Minkowski core'' contributes virtually nothing at high density; in the infinite past of the Hayward cosmology, by contrast, the intrinsic ``dS core'' is superimposed on the extrinsic $\varLambda$, so that ${H_{\text{Hayward},\varLambda}}^2 \to \left(3/l^{2}+\varLambda\right)/3$, implying a larger effective cosmological constant $\bar{\varLambda}_{\text{eff}} = 3/l^2 + \varLambda$.

%----------------------------------------------------
\subsection{Robustness of the asymptotic Minkowski past}
\label{2.4} %"2.4 of the main body"
%----------------------------------------------------
The pure new cosmology ($\varLambda=0$) approaches Minkowski spacetime in the infinite past with $H \to 0$ and all curvature scalars vanishing.
From the perspective of dynamical systems, it is important to ascertain whether this asymptotic state is a stable attractor of the homogeneous dynamics, or whether it relies on extreme fine-tuning of initial conditions.

We address this question for a general mixture of dynamical matter fluids with an effective equation-of-state parameter $w$ (which may comprise dust, radiation, or other non-$\varLambda$ components).
The homogeneous dynamics is governed by the conservation law $\dot \rho  + 3H \rho \left( {1 + w} \right) = 0$ (Eq.~\eqref{CL-2}) together with the Friedmann equation ${H_{\text{MC}}}^2 = F_{\text{MC}}\!\left ( \rho \right ) = \left( 8\pi\rho/3 \right ) \exp\!\left( -4\pi\alpha_0\rho/3 \right )$ (Eqs.~\eqref{FE-U} and \eqref{F-rho}).
Excluding the trivial non-evolving case $w=-1$, the null energy condition requires $w > -1$.
Introducing the inverse-density variable $u \equiv 1/\rho$ maps the early-time Minkowski limit $\left ( \rho, H \right ) \to \left ( \infty, 0 \right )$ to the finite point $\left ( u, H \right ) = \left ( 0, 0 \right )$, and transforms the dynamical equation into:
\begin{equation}
\label{DE-u} %"Dynamical Equation in terms of u"
\dot{u} = 3\left ( 1+w \right )u\sqrt{F_{\text{MC}}\!\left ( u^{-1} \right )} = 3\left ( 1+w \right )u\sqrt{\frac{8\pi}{3u}\exp\!\left(-\frac{4\pi\alpha_0}{3u}\right)}\;.
\end{equation}

A \textit{standard} linear stability analysis---expanding around the fixed point and computing the Jacobian---is \textit{not} applicable here.
At $u=0$ the function $\psi\!\left ( u \right ) \equiv \dot{u}$ satisfies $\psi'\!\left ( 0 \right ) = 0$, so the Jacobian vanishes and linear stability analysis is inconclusive.
One might then attempt to examine higher-order Taylor terms---the standard remedy when linearization fails (e.g., for $\psi\!\left ( u \right )=u^3$ where $\psi'\!\left ( 0 \right ) = 0$ but $\psi'''\!\left ( 0 \right ) \neq 0$).
In the present case, however, \textit{all} derivatives $\psi^{(n)}\!\left ( 0 \right )$ vanish: the exponential factor $\exp\!\left(-4\pi\alpha_0/3u\right)$ forces the Taylor series of $\psi\!\left ( u \right )$ to be identically zero, yet $\psi\!\left ( u \right )>0$ for any $u>0$.
(This is the real-axis manifestation of an essential singularity at $u=0$---the function is smooth but non-analytic.)
No finite-order Taylor expansion can capture the dynamics near the fixed point.
For a one-dimensional system, however, the sign of $\dot{u}$ in the neighborhood of the fixed point alone fully determines stability, without any need to differentiate.

According to Eq.~\eqref{DE-u}, for any $w>-1$, $\dot{u} \to 0$ as $u\to 0^+$ owing to $\exp\!\left(-4\pi\alpha_0/3u\right)$, confirming that $u=0$ is indeed a fixed point of the dynamical system.
More importantly, for $u>0$, $\dot{u} > 0$, which indicates that in the forward direction $u=0$ is an unstable fixed point (repeller)---the system evolves away from it.
By time reversal, however, a forward repeller is precisely an asymptotically stable fixed point (i.e., attractor) in the backward direction---all trajectories converge to $u=0$ as $t \to -\infty$, meaning the asymptotic Minkowski past is approached regardless of initial data.
The robustness against fine-tuning is further reinforced by the one-dimensional character of the homogeneous dynamics: $H$ is a single-valued function of $u$ through the Friedmann equation, so every trajectory follows the same $H\!\left( u \right)$ curve in phase space; different initial conditions merely shift the system along this curve without altering its past asymptotic destination.

\begin{figure*}[ht]
\centering
\vspace{1em}
\includegraphics[width=0.98\textwidth]{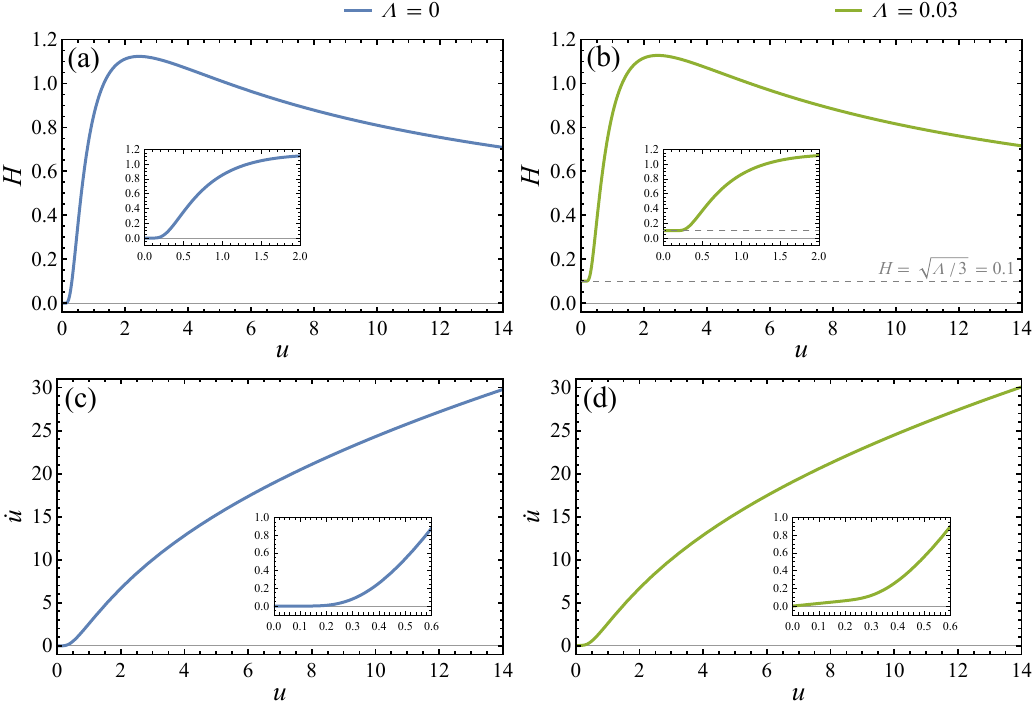}
\caption{Phase portraits of the new cosmology. (a) and (b): $H$--$u$ curves for the pure case  ($\varLambda=0$) and the $\varLambda$-extended case ($\varLambda=0.03$), respectively; (c) and (d): the corresponding $\dot{u}$--$u$ phase portraits in the dust-only scenario ($w=0$).}
\label{PhasePortraits}
\end{figure*}

To illustrate the above analytical results, the dust-only scenario ($w=0$) serves as a representative numerical example, with its phase portraits presented in Fig.~\ref{PhasePortraits} for both the pure case ($\varLambda=0$, left column) and the $\varLambda$-extended case ($\varLambda>0$, right column).
Both $H$ and $\dot{u}$ are single-valued functions of $u$, determined entirely by the corresponding Friedmann equation; all numerical trajectories trace these same curves regardless of initial data, indicating that the one-dimensional homogeneous dynamics admits a common past fixed point (global attractor) irrespective of initial conditions.

For the $\varLambda$-extended case, the Friedmann equation~\eqref{FE-MC-Lambda} gives ${H_{\text{MC},\varLambda}}^2 = F_{\text{MC}}\!\left ( \rho \right ) + \varLambda/3$.
As $\rho \to \infty$ ($u \to 0^+$), the exponential suppression drives $F_{\text{MC}}\!\left ( \rho \right ) \to 0$, leaving ${H_{\text{MC},\varLambda}}^2 \to \varLambda/3$.
The evolution equation for $u$ becomes:
\begin{equation}
\label{DE-u-Lambda} %"Dynamical Equation with Lambda in terms of u"
\dot{u} = 3\left ( 1+w \right )u\sqrt{F_{\text{MC}}\!\left ( u^{-1} \right ) + \frac{\varLambda}{3}}\;.
\end{equation}
For $u \to 0^+$, the square root approaches $\sqrt{\varLambda/3} > 0$, so $\dot{u} \propto u$ with a \textit{finite} coefficient $3(1+w)\sqrt{\varLambda/3}$, rather than the exponential suppression of the pure case---a difference clearly visible by comparing the $\dot{u}$--$u$ curves in panels (c) and (d) of Fig.~\ref{PhasePortraits}.
The $u=0$ boundary remains a past attractor---$\dot{u} \to 0$ as $u \to 0^+$ and $\dot{u} > 0$ for $u > 0$, the same sign structure as in the pure case---but the approach is exponential in $t$ ($u\!\left ( t \right ) \sim \exp\!\left( 3\left ( 1+w \right )\sqrt{\varLambda/3}\,t \right)$ as $t \to -\infty$) in contrast to the logarithmic approach in the pure case ($u\!\left ( t \right ) \sim \left ( 2\pi\alpha_0/3 \right )\!/\ln\!\left ( -t \right )$ as $t \to -\infty$, owing to the extreme flatness of $\dot{u}\!\left ( u \right )$ near $u=0$).
Thus, while the pure new cosmology has an asymptotic Minkowski past attractor ($H_{\text{MC}} \to 0$, as $u \to 0^+$) with logarithmic convergence, the $\varLambda$-extended version has an asymptotic dS past attractor ($H_{\text{MC},\varLambda} \to \sqrt{\varLambda/3}$, as $u \to 0^+$) with exponential convergence---the distinct asymptotic values of $H$ being directly visible in panels (a) and (b) of Fig.~\ref{PhasePortraits}.

Taken together, in both the pure and the $\varLambda$-extended cases, the past attractor is robust---no extreme fine-tuning of initial conditions is required---but the asymptotic Hubble parameter and the convergence rate differ qualitatively.

%====================================================
\section{Summary and discussion}
\label{conclusion}
%----------------------------------------------------
In summary, we have proposed a new non-singular cosmological model matching a Minkowski-core regular BH model \cite{Ling:2021olm} based on the modified OS framework, and investigated its dynamics in detail.

The comparative studies in both dust-only and scalar-field scenarios have been conducted with those of the cosmology corresponding to the Hayward BH model (featuring an asymptotically dS core), LQC, and the classical standard cosmology.
The results indicate that the three non-singular cosmological models, which share the identical late-time asymptotic behavior, differ markedly in the early-time regime.
Specifically, the big-bang singularity in GR is replaced with a big bounce, an asymptotically dS epoch, and an asymptotically Minkowski epoch in LQC, the Hayward case, and the new model, respectively, in excellent agreement with the characteristics near the center of their corresponding BH models.
Additionally, the investigation of the scalar-field dynamics demonstrates that inflation can be naturally embedded within each of these three cosmological frameworks beyond GR.

We now focus on our newly proposed cosmological model.
As illustrated collectively by the evolutions of the scale factor and the Hubble parameter in the dust-only scenario (Fig.~\ref{H&a}(a)--(d)), the curvature scalars as functions of the scale factor (Fig.~\ref{R&K}(a) and (c)), and the phase trajectories of the scalar-field dynamics (Fig.~\ref{Dyn}(d) and (d$'$)), the new cosmology asymptotically approaches a flat Minkowski spacetime in both the very-early-time (small-scale, high-density) and the very-late-time (large-scale, low-density) regimes.
In particular, the Hubble parameter as well as the curvature scalars tend to vanish in the limits of both the infinite past and the infinite future, and the scalar-field dynamics displays a nearly symmetric oscillatory behavior in these two regimes.
This is rather intriguing---within this model, the universe exhibits very similar geometric features in the remote past and remote future.
Moreover, unlike in classical cosmology, such a universe does not originate from a finite-time initial singularity but instead evolves smoothly from the infinite past up to the present epoch (as is also the case in the other two non-singular cosmologies; indeed, this appears to be a generic feature of all non-singular cosmologies).
Although at first sight such a picture may appear counterintuitive, it becomes less surprising once one recalls that our universe is also quite plausibly expected not to end at any finite future time.

Focusing specifically on the new model, we have also explored the scenarios of other cosmological fluids (such as radiation and dark energy), as well as the more realistic and general scenario of multiple coexisting fluids, in particular incorporating a positive cosmological constant.
The results are essentially consistent with the preceding analysis, except in the cases involving a cosmological constant, where the symmetry between the early- and late-time regimes shifts from asymptotically Minkowski to asymptotically dS.
Furthermore, we have also assessed the robustness of its asymptotic past via a dynamical stability analysis.
The asymptotic-past state proves to be a robust attractor in both the pure and the $\varLambda$-extended versions of the model, with no need to fine-tune the initial conditions, while the two cases differ qualitatively in the asymptotic Hubble parameter and the rate of approach to the attractor.

It merits further explicit discussion of the possible observational signatures of this model in the early universe, which would help distinguish it from the LQC and Hayward alternatives.
The four models exhibit qualitatively distinct very-early-time behaviors (for $\varLambda = 0$, namely, a big-bang singularity for GR, a quantum big bounce for LQC, an asymptotic dS epoch for the Hayward cosmology, and an asymptotic Minkowski epoch for the new model), which would set different initial conditions for the subsequent inflationary phase and could not merely be of theoretical interest but also leave observable imprints in principle.
Regarding the cosmic microwave background (CMB) power spectrum, the quasi-Minkowski pre-inflationary phase of the new cosmology ($H_{\text{MC}} \approx 0$ for $\varLambda = 0$) would naturally tend to suppress the amplitude of large-scale pre-inflationary perturbations, potentially leading to a power deficit at the largest angular scales, in contrast to LQC and the Hayward cosmology, which retain non-zero and finite pre-inflationary Hubble parameters and would predict different infrared features.
Among the three non-singular models, LQC is currently the only one for which quantitative CMB predictions have been developed: full Markov Chain Monte Carlo (MCMC) analyses with Planck 2018 data show that the dressed-metric LQC power spectrum can simultaneously alleviate the large-scale power anomaly and the lensing amplitude tension, without phenomenological adjustments put in by hand just before or during the slow roll \cite{Ashtekar:2020tension,Ashtekar:2021tango}; a recent Bayesian analysis using the non-oscillatory vacuum in hybrid LQC yields consistent conclusions \cite{Martin-Benito:2023nky}.
Comparable predictions for both the Hayward and the new cosmologies remain to be developed.
Concerning primordial gravitational waves, the near-vanishing Hubble parameter of the new cosmology at early times would tend to suppress tensor-mode production during the pre-inflationary era, in contrast to the LQC and Hayward cases.
Once the scalar field enters the slow-roll regime, however, all three models support a standard inflationary attractor at the background level (as demonstrated in subsection~\ref{2.2}), so the tensor spectrum on smaller scales (inflationary) would be similar across the three models.
The most promising model-specific discriminator thus lies in the \textit{pre-inflationary} contribution to the tensor spectrum at the very largest scales (lowest frequencies).
Ultimately, while a quantitative evaluation of CMB and primordial gravitational-wave signatures requires a full treatment of linear perturbations on the background of the new cosmology---a prerequisite beyond the scope of the present homogeneous analysis and deferred to future work---the qualitative differences identified here provide clear, physically motivated directions for distinguishing the three non-singular cosmologies through early-universe observations.

%====================================================
\acknowledgments
This work was partially supported by NSFC, China (Grant Nos.~12275166, 12311540141 and 12375055).

%====================================================
%附录
\appendix
%====================================================
\section{Interior-exterior matching via the junction conditions}
\label{AppendixA}
%----------------------------------------------------
In this appendix, we present a detailed derivation of the relations between the interior region ${{\cal M}^-}$ and the exterior region ${{\cal M}^+}$ in the Oppenheimer--Snyder (OS) dust collapse model discussed in subsection~\ref{1.1}, obtained by matching these two regions along the interface $\varSigma$ through the Darmois--Israel (DI) junction conditions.

The dust ball surface $\varSigma$ can be characterized by $\left ( t, {r_0}, \theta, \varphi \right )$, and then the induced metric on this surface, according to the interior metric \eqref{IM}, can be straightforwardly given by:
\begin{equation}
\label{I-IndM} %"the Interior Induced Metric"
\left.\mathrm{d} s_{-}^{2} \right |_{\varSigma }= - {\mathrm{d}t}^2 + {a\!\left( t \right)}^2 {{r_0}^2}{\mathrm{d}\Omega}^2 \;.
\end{equation}
When observed from the outside, the surface $\varSigma$ can be described as $\left ( {{\tilde t}\!\left ( t \right )}, {{\tilde r}\!\left ( t \right )}, \theta, \varphi \right )$, where the exterior coordinates $\tilde t$ and $\tilde r$ are parametrized by $t$.
Then the induced metric on the outer surface following the exterior metric \eqref{EM} can be expressed as:
\begin{equation}
\label{E-IndM} %"the Exterior Induced Metric"
{\left. {\mathrm{d}s_{+}^2} \right|_\varSigma } = - \left( {{f_1}{\dot {\tilde t}^{\,2}} - {{f_2}^{-1}}{\dot {\tilde r}^2}} \right){\mathrm{d}t}^2 + {{\tilde r}\!\left ( t \right )}^2 {\mathrm{d}\Omega}^2 \;,
\end{equation}
where the overdot denotes the derivative with respect to $t$.
Matching the above inner and outer induced metrics, ${\left. {\mathrm{d}s_- ^2} \right|_\varSigma } = {\left. {\mathrm{d}s_+ ^2} \right|_\varSigma }$, results in the following equations:
\begin{gather}
1 = {{f_1}{\dot {\tilde t}^{\,2}} - {{f_2}^{-1}}{\dot {\tilde r}^2}}\;,
\label{MatchingIndMs1} %"the 1st eq. obtained by Matching Induced Metrics"
\\
a\!\left( t \right) {r_0} = {{\tilde r}\!\left ( t \right )}\;.
\label{MatchingIndMs2} %"the 2nd eq. obtained by Matching Induced Metrics"
\end{gather}

The world lines of the dust particles in ${{\cal M} ^-}$ including the boundary surface $\varSigma$, are the geodesics whose tangent vector is $ {\left ( {\partial}/{\partial t} \right )}^\mu $.
Moreover, $ {\left ( {\partial}/{\partial {\tilde t}} \right )}^\mu $ on ${{\cal M}^+}$ is a Killing vector field, implying the existence of a conserved energy $E$ along the geodesic tangent to $ {\left ( {\partial}/{\partial t} \right )}^\mu $ on $\varSigma$.
This energy is computed as:
\begin{equation}
\label{E} %"the conserved Energy"
E:= - {g^+_{\mu \nu}}{{\left( {\frac{\partial }{{\partial {\tilde t}}}} \right)}^\mu}{{\left( {\frac{\partial }{{\partial t }}} \right)}^\nu} = - {g^+_{{\tilde t}{\tilde t}}}{{\left( {\frac{\partial }{{\partial t }}} \right)}^{\tilde t}} = {f_1} \dot{\tilde t} \;,
\end{equation}
and it remains constant, i.e., $\dot E = 0$.
Combining Eq.~\eqref{MatchingIndMs1} with Eq.~\eqref{E} yields the following results:
\begin{align}
\label{tildetDot} %"the expression of tilde t Dot"
\dot{\tilde t} &= E {{f_1}^{-1}}\;,\\
\label{tilderDot} %"the expression of tilde r Dot"
\dot{\tilde r} &= - \sqrt {{E^2}{{f_1}^{-1}}{f_2} - {f_2}} \;.
\end{align}

Next, let's turn our attention to the extrinsic curvatures of $\varSigma$.
Choosing the outgoing unit normal vector field and performing some calculations, we obtain the components of the interior extrinsic curvature $K_{\mu \nu}^{-}$ as follows:
\begin{equation}
\label{I-EC} %"the Interior Extrinsic Curvature"
K_{t t}^{-} = 0\;,\quad K_{\theta \theta}^{-} = \frac{K_{\varphi \varphi}^{-}}{{\sin ^{2}}\theta} = a\!\left( t \right) r_0 = {\tilde r}\!\left( t \right)\;,\quad \text{others} = 0\;;
\end{equation}
and the components of the exterior extrinsic curvature $K_{\mu \nu}^{+}$ are obtained as follows:
\begin{equation}
\label{E-EC} %"the Exterior Extrinsic Curvature"
K_{t t}^{+} = 0\;,\quad K_{\theta \theta}^{+}=\frac{K_{\varphi \varphi}^{+}}{{\sin ^{2}}\theta} = {\tilde r} E \sqrt{{f_1}^{-1} {f_2}}\;,\quad \text{others} = 0\;.
\end{equation}
Here Eq.~\eqref{MatchingIndMs2} has been used in Eq.~\eqref{I-EC}, and the derivation of Eq.~\eqref{E-EC} has utilized Eqs.~\eqref{tildetDot} and \eqref{tilderDot}, as well as the constancy of the conserved energy $E$.
Matching the interior and exterior extrinsic curvatures of $\varSigma$ such that $K_{\mu \nu}^{-} = K_{\mu \nu}^{+}$ yields:
\begin{equation}
\label{MatchingECs} %"Matching Extrinsic Curvatures"
1 = E \sqrt{{f_1}^{-1} {f_2}}\;.
\end{equation}

By combining Eq.~\eqref{tilderDot} with Eq.~\eqref{MatchingECs}, we obtain:
\begin{equation}
\label{tilderDot2} %"the 2nd expression of tilde r Dot"
{\dot{\tilde r}}^2 = 1 - f_2 \;.
\end{equation}
According to Eq.~\eqref{tilderDot2} and the derivative of Eq.~\eqref{MatchingIndMs2} with respect to $t$, we arrive at the expression for $f_2$ as:
\begin{equation}
	\label{f2} %"the expression of f_2"
	f_2 = 1 - H^2{\tilde r}^2 \;.
\end{equation}
In deriving the equation above, we have used the definition of the Hubble parameter, given by $ H\!\left ( t \right ) := {\dot a\!\left ( t \right )}/{a\!\left ( t \right )} $.
Moreover, Eq.~\eqref{MatchingECs} also indicates that the metric function $ f_1 \equiv -g^{+}_{{\tilde t}{\tilde t}} $ is equivalent to $ f_2 \equiv \left ( g^{+}_{{\tilde r}{\tilde r}} \right )^{-1} $ up to a constant factor $E^2$, i.e.,
\begin{equation}
\label{Rf1f2} %"the Relation between f_1 and f_2"
{f_1}\!\left( \tilde r \right) = E^2 {f_2}\!\left( \tilde r \right) \;.
\end{equation}
This means that, in the exterior metric \eqref{EM}, one can choose $ E = 1 $ without loss of generality, which is equivalent to absorbing a constant $E$ into the time coordinate $\tilde t$.
Upon doing so, we have:
\begin{equation}
\label{f1f2} %"the expression of f_1 and f_2"
{f_1}\!\left( \tilde r \right) = {f_2}\!\left( \tilde r \right) = 1 - H^2 {\tilde r}^2 \;.
\end{equation}

%====================================================
\section{Loop quantum cosmology and its matching black-hole model}
\label{AppendixB}
%----------------------------------------------------
%%%%%%%%%%%%%% LQG case %%%%%%%%%%%%%%
In the framework of loop quantum gravity (LQG), a quantum OS model applying the interior-to-exterior approach has been proposed in Refs.~\cite{Lewandowski:2022zce, Yang:2022btw}.

The loop quantum cosmology (LQC) model describes a non-singular universe, which resolves the big bang singularity in classical GR by replacing it with a non-singular big bounce at the effective level.
The effective Friedmann equation in LQC is given by \cite{Ashtekar:2006rx, Ashtekar:2006uz, Ashtekar:2006wn}:
\begin{equation}
\label{FE-LQG} %"the Friedmann Equation in LQC(G))"
{H_{{\text{LQG}}}}^2 = \frac{{8\pi}}{3}\rho \left( {1 - \frac{\rho }{{{\rho _{\text{c}}}}}} \right)\;,\quad \text{where} \quad {\rho_{\text{c}}} := \frac{3}{{8\pi {\gamma ^2}\varDelta }}\;.
\end{equation}
Here, the quantity $\varDelta := 4\sqrt{3}\pi\gamma {l_{\text{P}}}^2$, with the constant $\gamma > 0$ called the Barbero-Immirzi parameter \cite{BarberoG:1994eia, Immirzi:1996di} and the Planck length $l_{\text{P}} = 1$, serves as the LQC area gap, signifying the minimum non-zero eigenvalue of the quantum area operator.
And the Barbero-Immirzi parameter $\gamma$ has been determined to be $0.2375$ by the calculation of BH entropy \cite{Domagala:2004jt, Meissner:2004ju}.
In this context, ${\rho _{\text{c}}}$ is the critical density, representing the finite upper bound of the energy density $\rho$ at which the big bounce occurs.

The authors of Refs.~\cite{Lewandowski:2022zce, Yang:2022btw}, by utilizing the above effective Friedmann equation \eqref{FE-LQG} for the interior and the relation \eqref{f}, derived the exterior matching BH metric, where:
\begin{equation}
\label{BH-LQG} %"the metric function of BH in LQG"
f\!\left( \tilde r \right) = {f_{{\text{LQG}}}}\!\left( \tilde r \right) \equiv 1 - \frac{{2M}}{\tilde r} + \frac{{\alpha {M^2}}}{{{\tilde r}^4}} = 1 - \frac{{2M}}{\tilde r} \left ( 1 - \frac{{\alpha {M}}}{2{{\tilde r}^3}} \right ) \;,
\end{equation}
with $\alpha := 4{\gamma^2}\varDelta$.
From Eq.~\eqref{tilderDot2}, it can be seen that the collapsing velocity of the dust surface $\varSigma$ is $ v := \left | {\dot{\tilde r}} \right | = \sqrt{1 - f\!\left ( \tilde r \right )} $.
For the LQG case, $ v\!\left ( \tilde r \right )  = \sqrt{1 - f_{\text{LQG}}\!\left ( \tilde r \right )} = \sqrt{\frac{{2M}}{\tilde r} - \frac{{\alpha {M^2}}}{{{\tilde r}^4}}} = \sqrt{\frac{{{2M}}}{{{\tilde r}^4}}\left ( {\tilde r}^3 - \frac{\alpha M}{2} \right )} $ vanishes when the dust ball radius ${\tilde r} = {\left ( \frac{\alpha M}{2} \right )}^{\frac{1}{3}} \equiv {\tilde r}_{\text{b}}$.
At this point, the dust ball reaches the critical density ${\rho _{\text{c}}}$, stops collapsing, and bounces, and then transitions to an expanding phase.
Consequently, the bounce radius ${\tilde r}_{\text{b}} \equiv a_{\text{min}}r_0$ stands for the minimum radius of dust ball surface $\varSigma$, and the effective form \eqref{BH-LQG} of the exterior metric function is actually well-defined only over the dynamically accessible region of $\varSigma$, namely ${\tilde r} \in \left [{\tilde r}_{\text{b}},\infty \right )$.

% Asymptotic behavior of the two metrics.
In the classical (or equivalently, large-scale) limit where ${\tilde r} \equiv a\!\left( t \right) {r_0} \gg \sqrt[3]{8\sqrt{3}\pi{\gamma^3}M{l_{\text{P}}}^2}$ and thus $\rho \ll {\rho _{\text{c}}}$, both the effective Friedmann equation \eqref{FE-LQG} and the BH metric function \eqref{BH-LQG} reduce to their classical counterparts (Eqs.~\eqref{FE-GR} and \eqref{BH-GR}, respectively).

%====================================================
\section{Hayward black hole and its matching cosmological model}
\label{AppendixC}
%----------------------------------------------------
%%%%%%%%%%%%%% Hayward case %%%%%%%%%%%%%%
A modified OS model, based on the Hayward metric \cite{Hayward:2005gi} for the exterior geometry of the collapsing dust ball, has more recently been put forward in Ref.~\cite{Bobula:2024jlh}.

The metric function of the Hayward spacetime takes the following form:
\begin{equation}
\label{BH-Hayward} %"the metric function of BH in Hayward case"
f\!\left( \tilde r \right) = {f_{{\text{Hayward}}}}\!\left( \tilde r \right) \equiv 1 - \frac{{2M{{\tilde r}^2}}}{{{{\tilde r}^3} + 2{l^2}M}} \;,
\end{equation}
where $l$ is a constant expected to be the Planck length ${l_{\text{P}}} = 1$ or of the same order.
% about the BH
The Hayward BH is a typical example of the regular BH with an asymptotically de Sitter (dS) core which replaces the central singularity.
It can be observed that as ${\tilde r} \to 0$, ${f_{{\text{Hayward}}}}\!\left( \tilde r \right) \sim 1-{{\tilde r}^2}/l^2$, and the Einstein tensor $G_{\mu\nu} \sim -\left ( 3/l^{2} \right ){g^+_{\mu \nu}}$, indicating that this BH behaves as a dS spacetime with a cosmological constant $\varLambda_{\text{eff}} = 3/l^{2}$ at its center.
Moreover, its Ricci scalar $R_{\text{ext}}$ and Kretschmann scalar $K_{\text{ext}}$ are finite everywhere, particularly at the center $R_{\text{ext}} \to 12/l^{2}$ and $K_{\text{ext}} \to 24/l^{4}$.
On the other hand, in the classical or large-scale limit, namely ${\tilde r} \gg \sqrt[3]{2{l^2}M}$, ${f_{{\text{Hayward}}}}\!\left( \tilde r \right)$ recovers its classical counterpart ${f_{{\text{GR}}}}\!\left( \tilde r \right)$.
Furthermore, at large scales, ${f_{{\text{Hayward}}}}\!\left( \tilde r \right)$ behaves as:
\begin{equation}
\label{AB-BH-Hayward} %"the Asymptotic Behavior of BH in Hayward case"
{f_{{\text{Hayward}}}}\!\left( \tilde r \right) = 1 - \frac{2 M}{\tilde r} \left ( 1 - \frac{2 l^2 M}{{\tilde r}^3} + {\cal O}\!\left( {\left( \frac{2 l^2 M}{{\tilde r}^3} \right)\!}^2 \right )\ \right )\;,
\end{equation}
which is identical to ${f_{{\text{LQG}}}}\!\left( \tilde r \right)$ up to ${\cal O}\!\left( {\left( {2 l^2 M}/{{\tilde r}^3} \right)}^2 \right )$-terms if one identifies $2l^2$ with $\alpha/2$, i.e., takes $l = \sqrt{\alpha}/2 \equiv 2\sqrt{\sqrt{3}\pi\gamma^3}{l_{\text{P}}} \approx 0.5400{l_{\text{P}}}$.

By adopting the exterior-to-interior strategy via the relation \eqref{H2}, one employs Eq.~\eqref{BH-Hayward} as the exterior metric function and performs a simple calculation, which leads to the modified Friedmann equation governing the interior cosmology:
\begin{equation}
\label{FE-Hayward} %"the Friedmann Equation in Hayward case"
{H_{{\text{Hayward}}}}^2 = \frac{{8\pi}\rho }{{3 + 8\pi{l^2}\rho }}\;.
\end{equation}
The obtained modified cosmology turns out to be non-singular, exhibiting a dS era preceding (following) a power-law expansion (contraction), with a smooth transition between these two phases that ensures a graceful exit.
In parallel with the BH scenario, the Ricci scalar $R_{\text{int}}$ and Kretschmann scalar $K_{\text{int}}$ of this cosmological spacetime are finite all the time.
Interestingly, as $a \to 0$ they exactly coincide with those at the center of the Hayward BH, i.e., $\left. R_{\text{int}} \right |_{a\to 0} = \left. R_{\text{ext}} \right |_{{\tilde r}\to 0} = 12/l^{2}$ and $\left. K_{\text{int}} \right |_{a\to 0} = \left. K_{\text{ext}} \right |_{{\tilde r}\to 0} = 24/l^{4}$.
On the other hand, ${H_{{\text{Hayward}}}}^2$ also recovers its classical counterpart ${H_{{\text{GR}}}}^2$ in the classical limit with $a\!\left( t \right) {r_0} \gg \sqrt[3]{2{l^2}M}$ (and hence $\rho \ll {3}/{8\pi l^2}$).
Besides, in the large-scale, low-density regime, ${H_{{\text{Hayward}}}}^2$ can be expanded as follows:
\begin{equation}
\label{AB-FE-Hayward} %"the Asymptotic Behavior of the Friedmann Equation in Hayward case"
{H_{{\text{Hayward}}}}^2 = \frac{8\pi}{3}\rho \left ( 1 - \frac{8\pi l^2}{3}\rho + {\cal O}\!\left( {\left( \frac{8\pi l^2}{3}\rho \right)\!}^2 \right )\ \right )\;,
\end{equation}
which coincides with ${H_{{\text{LQG}}}}^2$ up to ${\cal O}\!\left( {\left( {8\pi l^2 \rho}/{3} \right)}^2 \right )$-terms upon taking $l = 2\sqrt{\sqrt{3}\pi\gamma^3}{l_{\text{P}}} \approx 0.5400{l_{\text{P}}}$.

%====================================================
%====================================================
\bibliographystyle{utphys}
\bibliography{Ref}
\end{document}